\documentclass[letterpaper,11pt]{article}
\usepackage[letterpaper, left=.8in, top=.8in, right=.8in, bottom=.8in,nohead,includefoot]{geometry}
\usepackage{color,charter,graphicx,natbib,enumitem,amsmath,amssymb,amsfonts,titlesec,placeins,bm,setspace} 
\usepackage[svgnames,x11names]{xcolor}
\usepackage[T1]{fontenc} 
\usepackage[pdftex]{hyperref}
\hypersetup{
	colorlinks,%
	linkcolor=MidnightBlue,  
	urlcolor=DarkSlateGray,   
	citecolor=RoyalBlue2  
	}
\titleformat*{\section}{\normalfont\Large\bfseries\blu}
\titleformat*{\subsection}{\normalfont\large\bfseries\blu}
\titleformat*{\subsubsection}{\normalfont\normalsize\bfseries\blu}
\def\blu{\color{RoyalBlue4}}           
\def\bone{\mathbf{1}}
\def\bzero{\mathbf{0}} 
\def\A{\mathbf{A}} 
\def\D{\mathbf{D}}

\def\V{\mathbf{V}}
\def\W{\mathbf{W}}  
\def\e{\textrm{e}}
\def\f{\mathbf{f}}  

\def\s{\mathbf{s}}  

\def\x{\mathbf{x}}
\def\y{\mathbf{y}}
\def\z{\mathbf{z}} 
\def\F{\mathbf{F}} 
\def\M{\mathbf{M}}
\def\R{\mathbf{R}}  
\def\C{\mathbf{C}} 
\def\V{\mathbf{V}} 
\def\W{\mathbf{W}} 

\def\bepsilon{{\bm\epsilon}}

\def\bnu{{\bm\nu}}
\def\bTheta{{\bm\Theta}}
 
\def\bOmega{{\bm\Omega}}

\def\btau{{\bm\tau}}

\def\cD{{\mathcal D}}\def\cM{{\mathcal M}}\def\cP{{\mathcal P}}\def\cY{{\mathcal Y}}

\newcommand{\RR}{\mathbb{R}}
\def\seq#1#2{#1{:}#2}
\def\eqn#1{eqn.~(\ref{eq:#1})}\def\Eqn#1{Eqn.~(\ref{eq:#1})}
\def\eqns#1#2{eqns.~(\ref{eq:#1},\,\ref{eq:#2})}

\def\bi{\begin{itemize}[noitemsep,topsep=3pt]}
\def\ei{\end{itemize}}\def\i{\item}
\def\bn{\begin{enumerate}[noitemsep,topsep=3pt]}
\def\en{\end{enumerate}}
\def\beq#1{\begin{equation}\label{eq:#1}}\def\eeq{\end{equation}}
\def\beas{\begin{eqnarray*}}\def\eeas{\end{eqnarray*}}
\def\bea{\begin{eqnarray}}\def\eea{\end{eqnarray}}

\def\starzero{\mathbf{m}_{\bzero}}
\def\starbm{\mathbf{m}}
\def\starm{m}

\def\portar{m}
 
0

\begin{document}

\title{\blu\bf\huge Bayesian Predictive Decision Synthesis}
\author{{\Large Emily Tallman \& Mike West}\\ 
{\em\blu emily.tallman@duke.edu, mike.west@duke.edu}\\ {} \\ 
Department of Statistical Science, Duke University \\ Durham NC 27708-0251, U.S.A.  }

\maketitle \thispagestyle{empty}
\begin{center}{\blu\Large\bf Abstract}\end{center}
 
Decision-guided perspectives on model uncertainty expand traditional statistical thinking about managing, comparing and combining inferences from sets of models. Bayesian predictive decision synthesis (BPDS) advances conceptual and theoretical foundations, and defines new methodology that explicitly integrates decision-analytic outcomes into the evaluation, comparison and potential combination of candidate models. BPDS extends recent theoretical and practical advances based on both Bayesian predictive synthesis and empirical goal-focused model uncertainty analysis. This is enabled by the development of a novel subjective Bayesian perspective on model weighting in predictive decision settings. Illustrations come from applied contexts including optimal design for regression prediction and sequential time series forecasting for financial portfolio decisions. 

\bigskip
\noindent{\blu\em Keywords:} 
Bayesian decision analysis, Decision-guided model weighting, Generalised Bayesian updating, Entropic tilting, Model uncertainty, Optimal design, Portfolio decisions

\section{Introduction}

Questions of model assessment, calibration, comparison and combination define continuing conceptual and practical challenges to all areas of quantitative modelling.   Recent developments in Bayesian thinking has  generated new  methodology motivated by a range of challenging applications
~\citep[e.g.][]{VanDijkEtAl2013,Amisano2017,Kapetanios2015,AastveitEtAl2019,McAlinnWest2018,West2020Akaike,VannitsemEtAl2021,AastveitEtAl2022}. However, while advancing  methodology for improved prediction,  such research developments have also raised foundational questions on the scope of model uncertainty analysis more broadly. 

A key theme is recognising explicit inferential goals to reflect the intended uses of models when addressing model assessment and uncertainty. This perspective is increasingly recognised~\citep[e.g.][and references therein]{ClydeIversen2013,McAlinnWest2018,McAlinnEtAl2019}, with numerous applied studies emphasising the relevance of differentially weighting models modulo specified goals~\citep[e.g.][]{Geweke2011,Geweke2012,NakajimaWest2013JBES,Kapetanios2015,West2020Akaike,McAlinn2021}.  This theme has been recently highlighted  in~\cite{LavineLindonWest2021avs} and~\cite{LoaizaMayaJOE2021}. These authors use explicit model weightings based on so-called \lq\lq Gibbs probabilities\rq\rq  that incorporate metrics related to utility (or \lq\lq score\rq\rq) functions evaluating predictions of specific, defined outcomes. Studies there highlight the relevance of this perspective, and its dominance over traditional \lq\lq goal neutral\rq\rq statistical approaches such as Bayesian model averaging when explicit goals can be defined and used to focus and guide the analysis.   
 
This paper contributes to this area by addressing foundational perspectives on model  comparison and combination that reflect  both predictive and decision goals. The perspective follows~\cite{Lindley1992} in emphasising both the inferential \lq\lq Yin\rq\rq and the decision analytic \lq\lq Yang\rq\rq of Bayesian analysis. 
Developments build on the foundation of Bayesian predictive synthesis (BPS:~\citealp{McAlinnEtAldiscussionBA2018,McAlinnWest2018,JohnsonWest2022}). Recent methodological developments linked to the BPS foundations are increasingly impacting forecasting and related applications, especially in economics and finance~\citep[e.g.][]{BassettiEtAl2018,McAlinnEtAl2019,AastveitEtAl2019,McAlinn2021,AastveitEtAl2022}.   A stylised class of BPS mixture models-- referred to as {\em mixture-BPS}-- is extended to integrate explicit decision goals and outcomes into the model uncertainty framework.  Modelling choices and implementation exploit 
entropic tilting (ET), originally introduced in predictive analyses to condition predictive distributions on sets of externally-imposed 
constraints~\citep{RobertsonET2005,KrugerET2017,West2021decisionconstraints}.  As developed here,    
ET defines a broader decision analysis setting for exploring, contrasting and integrating constraints into predictive inferences, with opportunities for exploitation in new ways.  Key foundational connections are made with generalised belief updating, in which data-based evidence is represented in likelihood functions constructed based on defined loss or utility functions~\citep{Tanner2008,BissiriWalker2016,bernaciak2022loss}
This links to integrating decision-analysis perspectives into Bayesian updating under model misspecification~\citep[e.g.][]{Zhang2006a, Zhang2006b,Tanner2008,WatsonHolmes2016}. These theoretical themes tie into Bayesian model scoring based on Gibbs model probabilities~\citep{LavineLindonWest2021avs,LoaizaMayaJOE2021} that explicitly focus on incorporating specific predictive goals into the model evaluation processes.

Integration of  ET with BPS defines  the new framework of {\em Bayesian predictive decision synthesis (BPDS)}.  This recognises that formal decision analysis will often yield differences across models in the implied optimal decisions and realised utilities, so a primary focus on decisions and their outcomes is relevant in  model comparison and combination.    Integration with other recent developments in decision-guided model scoring and weighting, including adaptive variable selection~\citep{LavineLindonWest2021avs}, enables the incorporation of information on historical predictions and decision outcomes  in the overall BPDS framework. This is especially relevant in development in sequential time series applications with dynamic models for forecasting and decisions. 

Section~\ref{sec:BPS} 
reviews aspects of the foundations and nature of {\em mixture-BPS} that is of central interest here. 
This is presented in the traditional predictive setting with no links to decision analysis. 
Section~\ref{sec:ETGB} 
discusses new, stylised versions of mixture-BPS models, and discusses theoretical foundations in both entropic tilting and generalised Bayesian updating. 
These new developments are again presented in the purely predictive setting, not addressing decisions. 
The extensions to integrate model-specific decisions, defining the new BPDS framework, are presented in
Section~\ref{sec:BPDS}. 
Detailed developments are made in two example settings. Section~\ref{sec:designexample} concerns optimal design variable selection in a simple regression setting as an illuminating, idea-fixing example. 
Section~\ref{sec:portfolioexample} develops an extensive example in sequential portfolio decision analysis in time series. Concluding comments are in Section~\ref{sec:conclude}.   
 
\section{Bayesian Predictive Synthesis (BPS) and Mixture Models \label{sec:BPS}}

\subsection{Traditional Model Uncertainty Framework \label{sec:BMA} } 

Begin with the usual model uncertainty  framework of a discrete mixture of distributions for outcomes of interest.   
The quantities of interest and notation are as follows. 
\bi
\i An outcome $q-$vector $\y$ in sample space $\cY$. Examples are the returns on $m$ equities at the next time period in a financial time series forecasting application or the outcome in a response surface design study. 
\i A set of $J$ models $\cM_j$, $j=\seq1J,$ such as a set of dynamic regression models for the return vector with different predictors across models, or other distinct model forms.
\i Model  $\cM_j$ defines a predictive density function $p_j(\y|\cM_j)$, the set of which is denoted by  $\cP  = \{ p_1(\y|\cM_1),\ldots,p_J(\y|\cM_J) \}.$
\i Model probabilities $\pi_j = Pr(\cM_j)$, such as from traditional Bayesian model averaging based on past data and information  denoted by $\cD.$ \i The dependence of models and model probabilities on $\cD$ is not made explicit in the notation, for clarity; such dependence is, of course, critical in applications and implicit throughout.
\ei
The predictive mixture $p(\y) = \sum_{j=\seq 1J}\pi_j p_j(\y|\cM_j)$ is implied under 
traditional Bayesian model averaging (BMA).  A well-known concern with BMA is that it does not address {\em model set incompleteness}, i.e., that all models $\cM_j$ may be poor predictors (as well as all \lq\lq wrong''), a key point that is revisited below in discussing BPS.

\subsection{Bayesian Predictive Synthesis: Mixture-BPS  \label{sec:BPSmix}} 

\subsubsection{Background and Mixture-BPS Model Form}
Bayesian predictive synthesis (BPS) adopts a subjective Bayesian view of model uncertainty in which the set of predictive densities $\cP$ is regarded by a Bayesian modeller as information to condition upon in forming their predictions. This is known as the \lq\lq supra-Bayesian'' perspective~\citep{LindleyEtAL1979}, and provides opportunities to adjust model-specific predictions for biases and miscalibration, and 
to at least partly address model set incompleteness.  BPS evolved from the theory of  agent opinion analysis~\citep{West1992c,West1992d} which built on foundations in~\cite{GenestSchervish1985}.  The approach is semi-parametric Bayesian, and results in identifying a subclass of predictive distributions that are consistent with the modeller's  partial specification of prior uncertainty about $\y$ and  $\cP.$   \cite{JohnsonWest2022}, and 
\cite{McAlinnWest2018},  discuss the foundational theory, and present a range of examples among which
{\em outcome dependent mixture-BPS} represents the special case of focus  here. This subclass of BPS models is the foundation for the 
decision-focused extensions in Section~\ref{sec:BPDS} below.  In one specific subset of mixture-BPS models, the modeller's predictive distribution-- conditional on learning $\cP$-- has the form
\begin{equation} \label{eq:BPSmixgeneral} 
f(\y|\cP) \propto \sum_{j=\seq 0J} \pi_j \alpha_j(\y) p_j(\y|\cM_j) 
\end{equation} 
where (i)  $\pi_{\seq0J}$ is a set of {\em initial model probabilities}; (ii)  $p_0(\y|\cM_0)$ is the density of a {\em baseline distribution} specified by the modeller and based  on a notional baseline model $\cM_0$; and (iii)   $ \alpha_j(\cdot)$, $(j=0{:}J),$ are non-negative weight functions. 

\Eqn{BPSmixgeneral} can be rewritten as  
\begin{equation} \label{eq:BPSf} 
f(\y|\cP)  = \sum_{j=\seq 0J} \tilde\pi_j f_j(\y|\cM_j) \ \ \textrm{where} \ \   f_j(\y|\cM_j) = \alpha_j(\y)p_j(\y|\cM_j)/a_j, \,\, j=\seq0J, 
\end{equation} 
in which
$$a_j = \int_{\y} \alpha_j(\y)p_j(\y|\cM_j)d\y \ \ \textrm{and} \ \  \tilde\pi_j = k \pi_j a_j \ \ \textrm{with} \ \ 
k^{-1} = \sum_{j=\seq 0J} \pi_j a_j.$$   
Before discussing the details of these ingredients and their implications, the  constructive use of the above theoretical result is emphasised. 
That is,  {\em models} of the form of~\eqn{BPSf} can be explored for various choices of the weight functions $\alpha_j(\y);$ choosing these  directly defines the  predictive $f(\y|\cP)$. 
The current paper adopts this perspective. Note also that, in applications in sequential time series forecasting,  each of the $\y,  \pi_j, \alpha_j(\cdot), \cP$ and of course $\cD$ become time-indexed; see examples in~\cite{JohnsonWest2022}. Our time series example in  Section~\ref{sec:portfolioexample} follows this path.   
  
\subsubsection{Mixture-BPS Weight Functions}
  
The $\alpha_j(\y)$ terms explicitly depend on the future outcome $\y$,  and can be exploited to address anticipated biases and lack of calibration in each $\cM_j$, as well as to increase/decrease weight on any model in different regions of the outcome space of $\y.$  For example, a specific model $\cM_i$ may be expected to predict relatively more accurately in regions where element $y_1$ is high, but poorly when $y_1$ is low.   This relates to considerations of relative areas of \lq\lq expertise'' across the model set. Appropriately chosen weight (or \lq\lq kernel'') functions $\alpha_j(\cdot)$ will then change the contributions of the $\cM_j$ to $f(\y|\cP)$  accordingly as 
they modify the $p_j(\y|\cM_j)$ to resulting $f_j(\y|\cM_j)$ densities. Correspondingly, the $\tilde\pi_j$ are modified model probabilities, adjusting the initial $\pi_j$ based on concordance of $\alpha_j(\y)$ with $p_j(\y|\cM_j)$. 
The consideration of outcome-dependent calibration led to
important early work using forms similar to~\eqn{BPSf} in empirical studies  in macroeconomic forecasting, beginning with~\cite{Kapetanios2015}. Mixture-BPS underpins such developments and extends such approaches to admit a possible baseline model indexed by $j=0$ as well as specified initial model  probabilities $\pi_j$ based on historical data and information~$\cD.$ 

\subsubsection{Baseline Distribution}
 
The  baseline distribution is an important feature; it explicitly admits and addresses the potential that \lq\lq all models are wrong\rq\rq.  That is, the question of model set incompleteness~\citep{McAlinnEtAldiscussionBA2018,AastveitEtAl2019,McAlinnEtAl2019,GiannoneEtAl2021},  earlier and often referred to as the~\lq\lq model space open\rq\rq-- or $\lq\lq\cM-$open\rq\rq-- 
setting~(\citealp{BernardoSmith1994};~\citealp[][section 12.2]{WestHarrisonYellowBook19972ndEdition};  ~\citealp{ClydeGeorge2004};~\citealp{ClydeIversen2013}). 
It is often important to adopt this view,  admitting that predictions based on any of the candidate models may be poor, and assign some non-zero probability on 
a \lq\lq safe-haven'' or fall-back predictive; see also~\cite{DeGroot1980} for a pertinent historical perspective. 
$p_0(\cdot|\cM_0)$ may be chosen, for example, as an over-dispersed density, putting higher mass on regions of $\y$ likely to be less well-supported under any of the $\cM_j.$   
Such a baseline will receive increasing weight as outcome $\y$ predictions from the model set of $J$ models are more and more inaccurate. 
Thus  larger values of $\tilde\pi_0$ will indicate that $\cM_j$ models for $j=\seq 1J$ are collectively performing poorly. The conceptual and technical relationships with the construction of baseline \lq\lq alternative models'' in Bayesian analysis in other areas is to be noted~\citep[][section 11.4]{West1986a,WestHarrisonYellowBook19972ndEdition}.

\subsubsection{\blu Initial Model Weights} 
 
The $\pi_j$ can be regarded as initial model weights based on historical data and prior information $\cD$ that the modeller deems relevant to forecasting $\y.$  In a sequential time series context, for example, they may be discounted versions of past BMA weights, as used, for example in 
~\cite{ZhaoXieWest2016ASMBI} and adaptive variable selection (AVS) in \cite{LavineLindonWest2021avs}.  More generally, the modeller has the flexibility to choose the initial weights as a separate consideration to that of specifying the $\alpha_j(\cdot) $ weight functions which address questions of anticipated model bias and relative areas of expertise. 
 
\subsubsection{\blu Example: BMA as a Special Case}  Based on the initial model probabilities $\pi_{\seq1J},$ it is immediate that 
BPS specialises to BMA with the choices
$\pi_0=0$ and $\alpha_j(\z_j)=1$ for $j=\seq 1J$.   Obviously, there is no future outcome dependency here and the input predictives $p_j(\y|\cM_j)$ are not modified prior to averaging. 

BMA does not recognise model set incompleteness. However,  this simple mixture-BPS example with a non-zero $\pi_0$ opens a path to potentially raise awareness of, and adapt to, the lack of predictive ability of the set of models. Such a minor extension of BMA does, of course, require 
an appropriately chosen $p_0(\cdot|\cM_0)$.  
 
\subsubsection{\blu Examples: AVS and Goal-Focused Scoring of Historical Predictions}  
Important special cases-- again with no outcome dependence--  link mixture weights to historical data and {\em past} outcomes in the predictive performance of each model. Scoring models modulo specified forecasting goals has been stressed in various ways~\citep[e.g.][]{Geweke2011,Geweke2012,NakajimaWest2013JBES,Kapetanios2015,West2020Akaike,McAlinn2021}, and recently highlighted  in AVS~\citep{LavineLindonWest2021avs} and targeted prediction~\citep{LoaizaMayaJOE2021}. 
 
Recall that $\cD$ denotes all prior information the modeller deems relevant to combining forecast information for $\y,$ recognizing that the initial model probabilities $\pi_j$ are inherently dependent on $\cD.$   In a sequential time series context, and in which models are distinguished based on chosen sets of covariates in a dynamic linear modelling framework, the AVS approach of~\cite{LavineLindonWest2021avs} uses {\em scores} that reflect past predictive accuracy on outcomes of applied interest. The term {\em adaptive variable selection} explicitly relates to the models being based on different sets of covariates, and the changes over time in chosen covariates in a sequential time series setting.  Examples of scores include predictions of multi-step ahead outcomes in the time series, and multi-step paths over several time periods ahead. Ignoring the time indexing for a sequential time series setting,  AVS uses  outcome-independent choices $\pi_0=0$ and $\alpha_j(\y) \equiv \alpha_j\propto  \exp\{\tau s_j(\cD)\}$ for $j=\seq 1J,$ not depending on $\y.$  Here 
$\tau$ is a chosen scale factor and $s_j(\cD)$ is a chosen univariate score for model $\cM_j$ based on historical information $\cD.$  This is a special case of mixture-BPS with dynamic extension for the time series setting. Then, note that the approach 
in~\cite{LoaizaMayaJOE2021} adopts a functional form of weights similar to those defining  AVS.

\section{Entropic Tilting and Generalised Bayes  \label{sec:ETGB}}

\subsection{Exponential Score Weight Functions \label{sec:EXPSCORE} } 

Adopt models in which the BPS weight functions are defined by
\beq{alphajBPS}  \alpha_j(\y) = \exp\{ \btau'\s_j(\y)\}, \,\, j=\seq0J,\eeq 
where each $\s_j(\y)$ is a $k-$vector of prescribed {\em scores} that depend on the future outcome $\y$.  The scores are chosen to reflect utilities, so that higher scores are desirable.  The use of a vector score allows consideration of multiple aspects of model evaluation, i.e., multi-attribute utilities.   
The $k-$vector $\btau$ defines relative weights and directional relevance of the elements of the score vector, so that the $\cM_j$  are relatively weighted according to the resulting aggregate score. 

The exponential, weighted-score form of $\alpha_j(\y)$ is underpinned by theoretical foundations presented in Sections~\ref{sec:ET} and~\ref{sec:GB} below.  It also parallels and extends score-based methods that use historical outcomes to define weight functions via so-called \lq\lq Gibbs model probabilities\rq\rq  in generalised Bayesian updating, AVS and allied approaches~\citep[][and references therein]{BissiriWalker2016,LavineLindonWest2021avs,LoaizaMayaJOE2021}.  Key innovations here are that model weights in~\eqn{alphajBPS}  depend  on (i) the as yet unobserved outcome $\y$, and (ii)  multiple predictive metrics in the vector score.  This underlies the extension to integrate decision outcomes in Section~\ref{sec:BPDS} below. Prior to that development, the underlying foundational theory is summarised. 

\subsection{Entropic and/or Exponential Tilting \label{sec:ET}} 

Using \eqn{alphajBPS}, the BPS predictive densities $f_j(\cdot)$ in \eqn{BPSf}  
are \lq\lq exponentially tilted\rq\rq modifications of the $p_j(\cdot|\cdot), $ i.e., $f_j(\y|\cM_j)\propto \exp\{ \btau'\s_j(\y)\} p_j(\y|\cM_j).$  Exponential/entropic tilting~\citep{RobertsonET2005,KrugerET2017,West2021decisionconstraints} provides  conceptual and theoretical foundation for this choice of the 
$\alpha_j(\cdot)$ as well as a constructive approach specifying the $\btau$ vector.  

First, recognise that the initial mixture model is fundamentally a {\em joint distribution} over outcomes and  models together, i.e. the joint density/mass function $p(\y,\cM_j) = p(\cM_j)  p(\y|\cM_j) = \pi_j p_j(\y|\cM_j)$  over $\y\in\cY, j=\seq1J.$  Extend this to include a baseline case $j=0$ defined by the modeller.  Then, for score vectors $\s_j(\y)$ of chosen functional forms in $\y$, this extended mixture implies the {\em initial expected score} $\starzero = E_p[\s_j(\y)] = \sum_{j=\seq0J} \pi_j\starbm_{j,\bzero}$ where $\starbm_{j,\bzero} = \int_{\y} \s_j(\y)p_j(\y|\cM_j)d\y$ is  the expected score under $p_j(\cdot|\cM_j).$   Since the score is a vector of utilities, $\starzero$ represents a vector of {\em benchmark} expected utilities-- benchmark as they are based on the initial distribution over $(\y,\cM_j)$.
 
Entropic tilting (ET) investigates modifications of this joint distribution that are consistent with different expected scores. The relevant thinking here is that of improving, i.e., increasing scores relative to the initial mixture.   Choose a $k-$vector $\starbm$ to define a {\em target expected score}; for example, elements of $\starbm$ may represent small \% increases in expected score relative to  $\starzero.$  Predictive distributions over $(\y,\cM_j)$
that have this or higher expected scores are then of interest-- simply as they {\em may} result in increased realised scores.  Exploring small/modest 
increases in expected scores relative to $\starzero$ reflects a perspective that, while the initial mixture has been defined based on analysis to date--  including past decision outcomes as well as predictions-- it still represents a specific, chosen model, and small perturbations of it may yield improved predictions and decisions. It is important that chosen score functions be interpretable, so that the numerical values of elements of 
expected scores can be understood to aid specification of $\starbm.$ As noted above, in many cases this will be naturally assessed using percent changes over the expected score under the initial mixture. Further discussion in context is given in the design and portfolio examples below.

In its original form,  ET  operationalises these ideas by considering  {\em all}   distributions 
$f(\y,\cM_j)$ under which $E_f[\s_j(\y)] =\starbm$; then, among such distributions, ET identifies  $f(\cdot,\cdot)$  that is 
K\"ullback-Leibler (KL) closest to $p(\cdot,\cdot).$ The natural KL direction is used:  the divergence is that of $p(\cdot,\cdot)$ from $f(\cdot), $ i.e., 
$KL(f{\vert\vert}p)   = E_f[ \log\{f(\y,\cM_j)/p(\y,\cM_j)\}]$ where the expectation is with respect to $f(\y,\cM_j).$   
While this does not, a priori, impose any other assumptions on the form of $f(\y,\cM_j),$   the resulting 
the optimisation theoretically  yields the unique solution $\tilde\pi_j f_j(\y|\cM_j) \propto \pi_j\alpha_j(\y)p_j(\y|\cM_j)$ where $\alpha_j(\y) $ has precisely 
the exponentially tilted form of~\eqn{alphajBPS}. The unique optimising vector $\btau$ depends on $\starbm$ and is implicitly defined so that the resulting BPS distribution  $\tilde\pi_j f_j(\y|\cM_j)$  over $(\y,\cM_j)$ has this expected score.  Generally, there is a one-one correspondence between $\starbm$ and $\btau$ emerging from convexity properties of exponential families~(see, for example, the supplementary material in~\citealp{TallmanWestET2022}). 

In addition to this  theoretical foundation for the choice of weighting functions  of \eqn{alphajBPS}, ET is constructive. It allows for calibration of mixture-BPS  by defining the vector $\btau$ consistent with a chosen target score $\starbm.$ 
Assuming the elements of the score vector are on contextually interpretable scales, specifying $\starbm$ is natural.  
Then, evidently the terms $a_j$ and hence $k$ in~\eqn{BPSf} depend on $\btau$,  and the resulting implicit equation to solve for $\btau$ reduces to 
\beq{ETBPSfindtau}  \starbm = k   \sum_{j=\seq0J} \pi_j  \frac{\partial a_j }{\partial\btau}. \eeq
This is often amenable to an easy numerical solution to compute $\btau$ for a given $\starbm.$ It is also useful in reverse-- to  evaluate expected scores for any given values of  $\btau$.  

ET has a broader,  \lq\lq relaxed''  optimality property.   Take target score $\starbm\ge \starzero$ where the inequality is strict in at least one element of the vector. Then 
the ET optimal $\btau$ ensuring expected score $\starbm$ does, in fact, minimise the KL divergence over {\em all possible} target scores $\starbm_\ast\ge \starbm$.   This, and more detailed background and theoretical developments of entropic tilting, is shown in the supplementary manuscript of~\citet{TallmanWestET2022}. This supplement details the  theory of ET for conditioning predictive distributions on constraints, includes new results related to connections with exponential families, elaborates on the \lq\lq relaxed ET'' (RET) 
extension, and includes  examples of relevance more broadly and beyond the use in this paper.

\subsection{Generalised Bayesian Updating and Approximate Models \label{sec:GB}} 

A complementary theoretical foundation for the score form of BPS synthesis functions in \eqn{alphajBPS} comes from
generalised and robust Bayesian updating, earlier referenced~\citep[e.g.][and others]{BissiriWalker2016}.  In particular, there are intimate connections with the theme of Bayesian updating based on approximate models/priors as discussed in~\cite{WatsonHolmes2016}.  These authors work in terms of losses rather than utilities, and with only one loss metric; hence their results apply here with $\s_j(\y)$ becoming scalar $s_j(\y)$ and with $-s_j(\y)$ representing their  loss.       
 
In a general setting, \cite{WatsonHolmes2016} address uncertainty about a specified prior for a parameter or set of parameters.  Mapping to the current setting, the \lq\lq parameters'' are  $(\y,\cM_j)$ and the baseline-extended initial model defines a specified prior $p(\y,\cM_j)$  over $\y\in\cY, j=\seq0J.$  
\cite{WatsonHolmes2016} ask questions about comparing other distributions  
 $f(\cdot,\cdot)$ with $p(\cdot,\cdot)$ in terms of $KL(f{\vert\vert}p)$ as used in ET above. 
A translation of the result discussed following Theorem 4.1 in their paper is key here. Consider all possible priors $f(\cdot,\cdot)$ in the K\"ullback-Leibler neighbourhood defined by  $KL(f{\vert\vert}p) \le C$ for some chosen $C>0.$    That is, distributions $f(\cdot,\cdot)$ that are \lq\lq close to'' $p(\cdot,\cdot)$ in this KL sense with $C$ defining how close.   Across this set of priors, find that $f(\cdot,\cdot)$ that maximises the implied 
expected utility (negative loss) $E_f[s_j(\y)].$ The unique result is the exponentially tilted modification of the initial prior given by 
$f(\y,\cM_j) \propto \pi_j \exp\{ \tau_C s_j(\y)\} p_j(\y|\cM_j).$  Here $\tau_C\ge 0$ depends on $C$ and, as $C$ varies, is monotonically increasing in $C$.  

This result is the theoretical dual of that based on ET, providing a complementary foundational perspective.   The ET perspective has two distinctions, however.  First, some of the interest in BPS, and decision-focused extensions below, lies in multiple utilities related to predictive and decision goals. The ET foundation admits multivariate scores and underlies the general result in~\eqn{alphajBPS}.  The dual generalised Bayesian updating result in~\cite{WatsonHolmes2016} concerned a univariate utility, with no obvious/easy theoretical multi-utility extension.  
The second point relates to interpretation in specification.  With interpretable scores specified on understandable utility scales in an applied context, the ET approach is accessible in that a target  score $\starbm$ can be related to the benchmark value(s)  $\starzero$ under the initial prior.  In contrast, the specification of a practically relevant and interpretable $C$ to define a K\"ullback-Leibler neighbourhood in the complementary view is less natural and interpretable.

\subsection{Local Perturbations of Expected Scores} 
Exploring target expected scores that are small/modest perturbations of those under the initial mixture is emphasised. With $\starzero = E_p[\s_j(\y)]$ implied under $p(\y,\cM_j),$  take target 
$\starbm=\starzero+\bepsilon$ where the elements of $\bepsilon$ are zero or small and positive, representing modest increases in one or more of the utility dimensions relative to initial expectations.   This yields insights into, and and practical suggestions for, the choice of $\btau.$ 
Taylor series expansion of $\starbm$ in $\btau$ around $\btau=\bzero$ yields $\btau = \V_0^{-1} \bepsilon$ where $\V_0 = V_p[\s_j(\y)], $ the covariance matrix of the score vector under $p(\y,\cM_j)$.  This follows easily from exponential family theory as detailed in the supplementary manuscript~\citep[][section 3]{TallmanWestET2022}. 
As with the mean score $\starzero,$  the score covariance matrix $\V_0$ will typically be easily computed-- either analytically or via direct Monte Carlo sampling depending on the forms of the chosen  score functions.  On a theoretical point, this local perturbation approximation also leads to $C=\btau'\V_0\btau/2 = \bepsilon'\V_0^{-1}\bepsilon/2$ as the value of the KL divergence minimised under ET. This also links to the KL-neighbourhood \lq\lq radius'' in the generalised Bayesian updating setting, and highlights the role of ET in extending that approach to multivariate utilities.

\section{Bayesian Predictive Decision Synthesis (BPDS) \label{sec:BPDS}}
 
 \subsection{Explicit Decision Context}
 Now consider an explicit decision context in which the Bayesian modeller is to make a decision whose outcome will be known once $\y$ is observed. 
 In general, model predictions can depend on the decision, while utility functions can depend on the model. Then, given any model $\cM_j,$ standard Bayesian decision theory  applies, with the modeller choosing optimal decisions to maximise expected utility. The earlier, decision-free context and its notation is then extended as follows:
\bi
\i The decision variable is a $d-$vector $\x$. 
\i In model $\cM_j$ and conditional on any potential decision $\x,$ 
\bi 
\i  the predictive distribution is $p_j(\y|\x,\cM_j),$  
\i  the modeller specifies a utility function $u_j(\y,\x),$  and as a result  
\i  the model-specific optimal decision is $\x_j   = \textrm{arg max}_{\x}\int_{\y} u_j(\y,\x) p_j(\y|\x,\cM_j)d\y.$
\ei 
\ei
\smallskip\noindent{\bf\blu Design Examples.} In an optimisation/experimental design context,   $\x$ is a vector of control variables or design points to be chosen. Suppose past experimentation has led to $\x_0$ as the \lq\lq current'' or local/recent value, and $\y_0$ is a chosen or target outcome for the optimisation problem.  Specific utility functions might, for example, balance the expected proximity of the future outcome $\y$ to the target $\y_0$ with a corresponding measure of closeness of the optimising value of $\x$ to the current value $\x_0$. Here the predictive distribution critically depends on the decision, while issues of balancing relative scales of $\y$ and $\x$ lead to utility functions that are typically model dependent.

\medskip\noindent{\bf\blu Portfolio Examples.} In financial forecasting for portfolio analysis, $\y$ is a vector of future financial returns or prices of assets, and $\x$ is the vector of portfolio weights to be chosen. It is assumed that the  portfolio is small enough in value that it does not impact the market; thus  the predictive distributions  do not depend on $\x$, i.e., $p_j(\y|\x,\cM_j) = p_j(\y|\cM_j).$ Portfolio utility functions are often model-dependent, however, with $u_j(\y,\x)$ often very dependent on aspects of $p_j(\y|\cM_j).$   For example, a common portfolio strategy specifies a \lq\lq target'' level of expected portfolio return; a chosen target must be achievable under the predictive distribution $p_j(\y|\cM_j),$ and this can sometimes involve modifying targets-- and hence the utility function-- in ways specific to $\cM_j.$ In other cases, a common utility function $u_j(\y,\x) = u(\y,\x)$ is used.

\subsection{Decision-Guided Extension of Mixture-BPS}
 
Conditional on any candidate decision vector $\x,$  the mixture-BPS structure of~\eqn{BPSf} is simply modified to make the conditioning explicit.  This gives the decision-dependent predictive
\beq{BPDSf} 
f(\y|\x,\cP)  = \sum_{j=\seq 0J} \tilde\pi_j(\x) f_j(\y|\x,\cM_j) \ \ \textrm{where each}\ \   f_j(\y|\x,\cM_j) = \alpha_j(\y|\x)p_j(\y|\x,\cM_j)/a_j(\x), 
\eeq
in which
\beq{BPDSajk} a_j(\x) = \int_{\y} \alpha_j(\y|\x)p_j(\y|\x,\cM_j)d\y \ \ \textrm{and} \  \ \tilde\pi_j(\x) = k(\x) \pi_j a_j(\x)\ \textrm{with} \ \
k(\x)^{-1} = \sum_{j=\seq 0J} \pi_j a_j(\x).\eeq
The weights $\alpha_j(\y | \x)$ can now depend on all aspects of analysis under $\cM_j$. This dependence can now include aspects of the decision setting, including model-specific optimal decisions  $\x_j$.   This is implicit in the subscript-$j$ notation (used to maintain notational clarity) and is central to the development of BDPS. If each $\alpha_j(\y | \x)$ depends on the corresponding $\x_j$, this  dependence transfers to impact on the $a_j(\x)$, the resulting BPDS model probabilities $\tilde\pi_j(\x)$, and the reweighted, model-specific conditional densities $f_j(\y|\x,\cM_j).$   Hence  the relative model evaluation and combination may be influenced by the differences across models in the decision of interest as well as  their relative abilities in predicting $\y.$ 

The departure from traditional model uncertainty analysis is to be stressed. The focus is now on decision-guided reweighting of models that then underlie a final decision.  The BPDS mixture in \eqn{BPDSf} is not a traditional predictive distribution; it builds on past predictive and decision outcomes through the $\pi_j$ but now-- through appropriately chosen weight functions $\alpha_j(\y|\x)$-- explicitly targets potentially improved decision-making $(\x)$ as well as improved prediction $(\y)$.   The portfolio example setting~(Section~\ref{sec:portfolioexample}) aids in understanding the new perspective. One model may have generated superior portfolio returns over recent time periods, resulting in this model having a larger $\pi_j.$ However, this specific model also having superior decision-outcome performance in expectation in the coming time period suggests (some, but perhaps modest) further increased weight on this model to underlie the final portfolio decision $\x$ implied, compared to other models with similar past predictive ability but worse expected outcomes.

\subsection{Predictive Decision Scores\label{sec:BPDSalphaj}}

The foundational theory of Sections~\ref{sec:ET} and~\ref{sec:GB} extends immediately. Each model score function  now also potentially 
depends on the model-specific optimal decision $\x_j,$  so the notation is extended to make this explicit: the BPS score function $\s_j(\y)$ becomes $\s_j(\y,\x_j).$ 
Then~\eqn{alphajBPS} generalises as
\beq{alphajBPDS}  \alpha_j(\y|\x) = \exp\{ \btau(\x)'\s_j(\y,\x_j)\}, \,\, j=\seq0J,\eeq  
 where now $\btau(\x)$ in general depends on the decision variable $\x$.  This is most easily understood from the ET perspective, as follows.  The initial distribution over $(\y,\cM_j)$ now generally depends on $\x$ through $p_j(\y|\x,\cM_j)$, so that the implied 
the benchmark expected score given $\x$ is $E_p[\s_j(\y,\x_j)|\x] = \starzero(\x),$ varying with $\x.$ Hence any specified target score $\starbm$ must be referenced to this $\x-$dependent benchmark; this generates $\x-$dependence in the resulting tilting vector $\btau(\x).$ 
This dependence is central in areas such as optimal design for decisions, as in the example in Section~\ref{sec:designexample} below. In other applications where the $p_j(\y|\x,\cD_j)=p_j(\y|\cD_j)$ do not depend on $\x$, such as in the portfolio example of Section~\ref{sec:portfolioexample} below, 
$\btau(\x)=\btau$ is constant as in the original BPS.   

 The combination of the $\s_j(\cdot,\cdot)$ and the $\pi_j$ drives the balance across models;  the $\s_j(\cdot)$ are chosen to favour models expected to score more highly, but the $\pi_j$'s incorporation of past performance provides a balance  to mitigate favouring overly ambitious but unrealistic models.
Similar comments are relevant in design examples~(Section~\ref{sec:designexample}).  In application, the goal would be to relatively weight models through both the historical performance ($\pi_j)$ and for expected performance using appropriate scores that address trade-offs in both outcome $(\y)$ and decision $(\x)$ spaces. The simulation example focuses on the novel contribution of weighting through expected performance, leaving the $\pi_j$ as uniform. The two example sections below highlight these concepts underlying BPDS while demonstrating potential impact in key decision settings.

\subsection{Evaluation of Tilting Vectors\label{sec:comptaux}}

Computations typically involve  iterative numerical search over values of $\x$ to define the final optimal  decision vector. 
Each step involves computing  the $\x-$dependent  tilting vector $\btau(\x)$ that satisfies the constraint   
$\starbm=E_f[\s_j(\y,\x_j)|\x]$ for the chosen, constant target score $\starbm.$ From \eqn{BPDSf} this identity is
$$\starbm = k(\x) \sum_{j=\seq0J} \tilde\pi_j(\x) \int_{\y\in\RR^m} \s_j(\y,\x_j) f_j(\y|\x,\cM_j)d\y. $$ 
With the $\tilde\pi_j(\x)$ and $k(\x)$ as defined in \eqn{BPDSajk}, and the $\alpha_j(\y|\x)$ of \eqn{alphajBPDS},  this identify becomes 
\beq{ETfindtau}  \starbm= k(\x)   \sum_{j=\seq0J} \pi_j  \frac{\partial a_j(\x) }{\partial\btau(\x)}.\eeq
This simplifies to \eqn{ETBPSfindtau} in the decision-free setting of BPS, as well as in decision problems-- such as the portfolio example in Section~\ref{sec:portfolioexample}-- in which the $p_j(\y|\cdot)$ and $\alpha_j(\y|\cdot)$ do not depend on $\x.$

Solving \eqn{ETfindtau} for $\btau(\x)$ defines the BPDS distribution $f(\y|\x,\cP)$ at this current value of $\x$. Numerical methods to solve for $\btau(\x)$ will involve the first and, typically, second derivatives of each $a_j(\x)$ with respect to $\btau(\x).$  
From~\eqn{BPDSajk} and ~\eqn{alphajBPDS}, these are 
\beq{jderivs}\begin{aligned}
\frac{\partial a_j(\x)}{\partial\btau(\x) } 
 &= a_j(\x)  \int_{\y\in\RR^m} \s_j(\y, \x_j)  f_j(\y|\x, \cM_j) d\y,   \\
\frac{\partial^2 a_j(\x)}{\partial\btau(\x)\partial\btau(\x) '} 
 &=  a_j(\x)  \int_{\y\in\RR^m} \s_j(\y, \x_j)\s_j(\y,\x_j)'   f_j(\y|\x, \cM_j) d\y. 
\end{aligned}\eeq
Newton-Raphson solution is a first choice for numerical evaluation of the tilting vector, with the obvious initialisation 
$\btau(\x)=\bzero$ corresponding to no tilting.  The expectations in \eqn{jderivs} will typically be approximated via Monte Carlo using samples generated from each of the   $p_j(\y|\x, \cM_j)$; hence access to direct sampling from these distributions is important for application.

\subsection{Decision Synthesis \label{sec:finaldecision}}

Under \eqns{BPDSf}{alphajBPDS}, the decision maker now faces the choices of a utility function to evaluate the final decision $\x$ and utility-based consequences.  This is now standard Bayesian decision analysis, with freedom and context-dependence in choosing utility functions.  That said, a novel and natural class of utility functions that links to the conceptual basis of BPDS emerges and is used in the examples below. This BPDS {\em reference utility} specification may be adopted as a baseline for comparison with other choices, but often as the basis for final decision analysis. 

In parallel to the probabilistic perspective underlying BPS,  broaden the final decision perspective to address  model uncertainty by defining a class of utility functions $U(\y,\x,\cM_j)$:  these quantify the utilities of  adopting a final decision $\x$ if the outcome is $\y$ and if model $\cM_j$ were to be chosen for use in predicting $\y$ using $f_j(\y|\x,\cM_j).$     The {\em reference BPDS utility} function proposed is
\beq{DSutility} U(\y,\x,\cM_j) =  \btau(\x)'(\s_j(\y, \x) - \mathbf{b})\eeq
where $\mathbf{b}$ is the upper bound of $\s_j(\y, \x)$ across all $j, \x, \y$. As $\x$ varies,  $\btau(\x)$  down-weights potential choices of $\x$ that require more tilting from the initial mixture to achieve the specified target score $\starbm$, while still rewarding high-scoring decisions. 
Then, \eqn{DSutility} also defines a novel contribution to multi-objective decision analysis:  $\btau(\x)$ induces a weighted average of the univariate utilities in the $\s_j(\y,\x)$ to underlie final decisions. 
 
In many applications, simulation will be used to evaluate  predictions and decisions; that is, samples of $\y|\x$ from the $p_j(\cdot|\cdot)$ that can be used to evaluate functions of the BPDS mixture $f(\cdot|\cdot).$ Such samples provide the basis to explore the implied predictive distribution of utilities both assuming a specific model $\cM_j$ and then-- leading to final decisions-- under the BPDS mixture.  Then, the optimal Bayesian decision is given by 
$ \text{arg max}_{\x} \bar U(\x)$ where 
    \beq{DSoptimal}
	          \bar U(\x) = E [ U(\y,\x,\cM_j) | \x ] = \btau(\x)'  \sum_{j=\seq 0J} \tilde\pi_j(\x)  \int_{\y} \{\s_j(\y,\x) - \mathbf{b}\}  f_j(\y|\x,\cM_j)d\y.
    \eeq
Similar to \eqn{jderivs}, these integrals can be approximated using direct Monte Carlo with samples from $p_j(\cdot|\cdot)$.  Some analytic tractability will arise depending on the choice of score functions.
Examples below include Section~\ref{sec:designexample}, in which $\btau(\x)$ depends on the decision vector $\x,$ and 
Section~\ref{sec:portfolioexample}, in which it does not.  The main implications of dependence  on $\x$ are computational, as noted in those sections. 

Optimisation over \eqn{DSoptimal} typically involves numerical search,. Some applications may impose additional constraints on $\x$ to aid this optimisation, such as requiring positive elements or linear constraints; then, more general optimisation approaches can be required. The portfolio example below is such a case.

\section{Example: Experimental Design Decisions \label{sec:designexample}}

A first, idea-fixing example  comes from simple Bayesian optimisation. Take  $\y=y$ to be univariate and $\x=x$ to be a single control variable. Suppose that $y_0$ is a chosen target value for the optimisation problem, and $x_0$ is the current value of the control variable. The decision maker aims to choose a new $x$ to rerun the experiment.  For instance, the U.S. Federal Reserve aims to choose policy variables, such as the Federal Funds interest rate $(x)$, to influence future inflation outcomes $(y)$, obviously in a very simplified example setting here. In this example,  $x_0$ is the current interest rate, and $x$ is the new rate being set to target inflation level $y_0$.   The decision maker's utilities here will highly score outcomes $y$ that are close to the target $y_0,$ but will also down-weight choices of $x$ that are far from the current $x_0$ (i.e., the aim is to encourage inflation towards the target, but also to not \lq\lq rock-the-boat'' in terms of avoiding overly aggressive changes in central bank policy). 

A simulation example using simple linear regressions is illuminating. Synthetic past data is generated 
from $y_i = 0.7 +  1.2x_{i} - 0.9z_{i,1} + \epsilon_i$ where $\epsilon_i \sim N(0, v)$ with $v=0.09,$ and  
$x_{i}, z_{i,1}, z_{i,2} \sim U(0, 2)$. The $\epsilon_i,x_i,z_{i,\ast}$ are independent,  and the simulation generates $i = 1,\ldots,10$ samples. 
By design, $z_2$ is a nuisance variable, not impacting on outcomes $y_i.$ 
A set of $J=3$ models $\cM_j$ are normal linear regressions differing only in the choice of covariates: $x$ in $\cM_1$,  $(x,z_1)$ in $\cM_2$, and  $(x, z_1,z_2)$ in $\cM_3$.   Each $\cM_j$ has (true) residual variance $v=0.09$,  and linear predictor 
 $ \mu_j + \beta_j x$ where $\mu_j = \beta_{j,0} + \beta_{j,1}z_1+\beta_{j,2}z_2$   in which    $\beta_{1,1}=\beta_{1,2}=\beta_{2,2}=0$.  The non-zero $\beta_\ast$ parameters have independent, zero-mean normal priors with variances 
$V(\beta_{j,0})= v_yc$, $V(\beta_j) =v_yc/v_x$, $V(\beta_{j,1}) =v_yc/v_{z,1}$, $V(\beta_{j,2}) =v_yc/v_{z,2}$ with $c=9$ and where 
$v_x, v_y, v_1,v_2$ are the sample variances of $x,y, z_1,z_2$, respectively, in the synthetic training data. Standard normal conjugate analysis provides normal posteriors for regression parameters and hence, at any future chosen value of $x,z_1,z_2,$ results in normal predictive distributions for $y$ in each $\cM_j.$

For the BPDS analysis, the initial weights $\pi_j$ are taken as uniform and the  baseline p.d.f.  is chosen as $p_0(y|x,\cM_0) = N(\bar\mu+\bar\beta x,\bar{v}(x)/\delta)$, having the $\pi-$weighted mean over models $\bar\mu+\bar\beta x$   and a variance given by the variance $\bar{v}(x)$  of the $\pi-$weighted mixture of the $p_j(\cdot|\cdot)$ inflated with a discount factor $\delta = 0.135$ (c.f.,  baseline inflation discount factors in different, relevant  settings, in~\citealp{WestHarrisonYellowBook19972ndEdition}).   
 
\begin{figure}[t!]
    \centering
    \includegraphics[width=.6\textwidth]{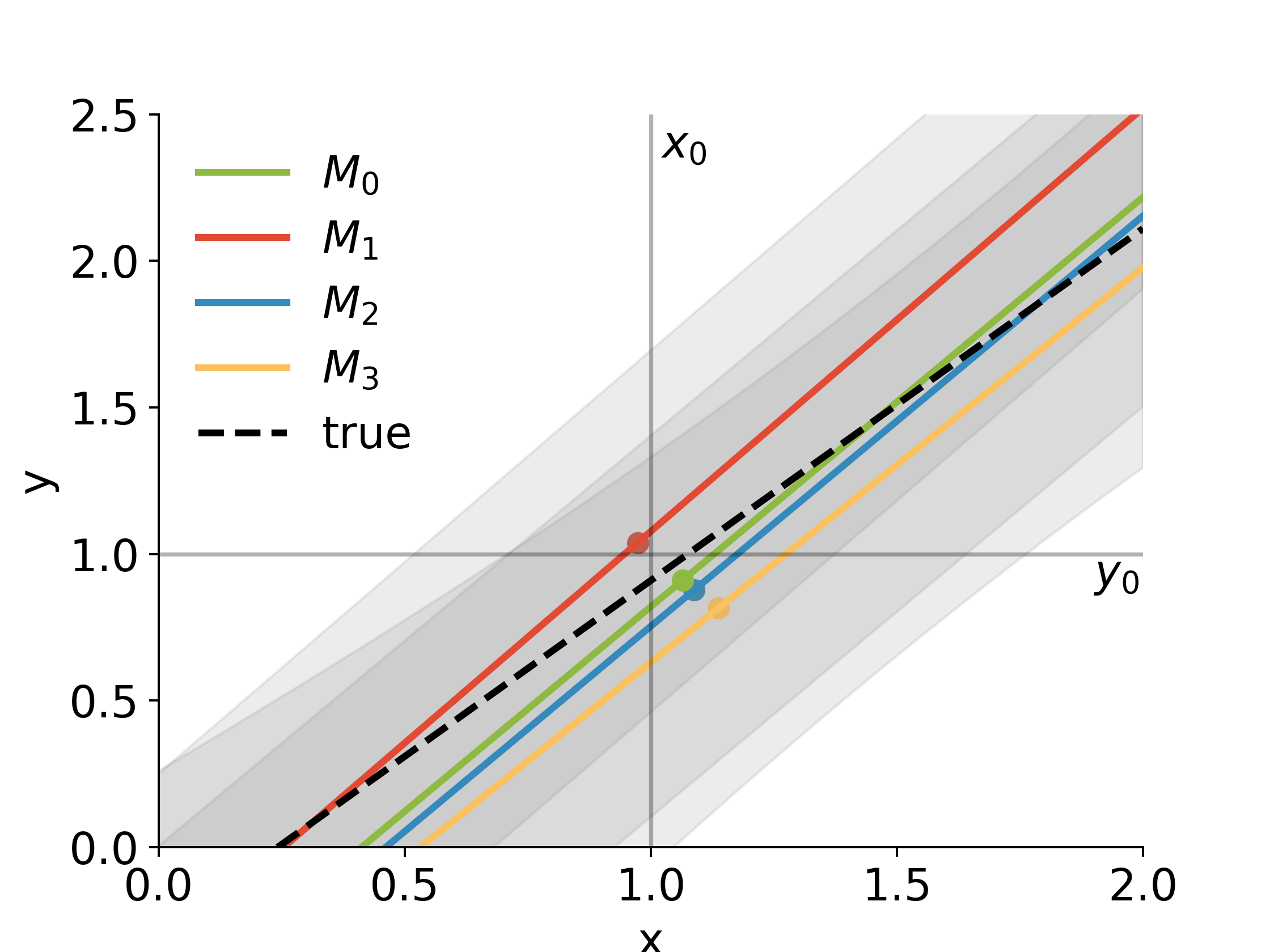}
    \caption{Optimal control example with $x_0=1$ and target $y_0=1$.  The figure shows the linear predictors with 95\% credible intervals from models $j=\seq13,$ and the linear predictor from the baseline model $j=0$ (no intervals shown for $j=0$ as they are wider than the plot bounds).  Model-specific optimal control points $x_j$ are indicated by dots.}
    \label{fig:ED_models}
\end{figure}

The design question is to choose $x$ with given settings  $z_1 = 1.1, z_2 = 0.1$, current control setting
$x_0=1$ and outcome target $y_0=1.$ Ignoring the dependence on the known $z_j$ in notation, for clarity, each predictive $p_j(y|x,\cM_j)$ is normal 
 at any chosen $x$.  Across a relevant range of $x,$  Figure~\ref{fig:ED_models} shows the true regression line, and the predictive regression with 95\% credible bands from each $\cM_j$ for $j=1,2,3,$ with bands for $\cM_0$ not shown due to its widely increased dispersion.
 Utility functions in each $\cM_j$  address interests in outcomes $y$ close to target $y_0$ while penalising design decisions $x$ that are far from the 
current control setting $x_0$. Specifically, 
$\cM_j$ adopts utility function $u_j(y,x) = -(y- y_0)^2/2 - c_j(x-x_0)^2/2$ with   $c_j = c b_j^2$ where $b_j$ is the posterior (to training data) mean of the $x-$coefficient $\beta_j$ in $\cM_j.$   This ensures that deviations of $x$ from $x_0$ are quantified on the same scale as those of $y$ from $y_0$. 
Since models have differing $b_j$ values, this is a setting of model-specific utility functions. Since all of the $b_j$ are similar in scale, with $b_0 = 1.443, b_1 =  1.399,  b_2 = 1.346, b_3 = 1.396,$, the utility functions are comparable.
The constant $c$ defines the balance of $y,x$ dimensions; the main example here is equally balanced, with $c=1$.   
The BPDS baseline utility $u_0(y,x)$  has the same form  with $c_0 = \bar c_j$, the $\pi-$weighted average of the $c_j.$

Taking expected utilities in each $\cM_j$ and solving for the optimal controls yields  the model-specific decisions
where $m_j$ is the posterior (to training data) mean of $\mu_j = \beta_{j,0} + \beta_{j,1}z_1+\beta_{j,2}z_2$ under $\cM_j$ at the given values of
 $z_1 = 1.1, z_2 = 0.1.$ 
These optimal control values are  indicated in Figure~\ref{fig:ED_models} in the numerical example.

BPDS adopts the univariate score functions $s_j(y, x) \equiv u_j(y,x_j) = -(y- y_0)^2/2 - c_j(x-x_0)^2/2 $ with $c_j=cb_j^2$, i.e., the same utilities as used in each model.  The following details are then implied. 
\bi
\i  In  model $\cM_j$, the weight functions $\alpha_j(y| x) = \exp\{\tau(x) s_j(y, x_j)\}$ are
$$\alpha_j(y| x) = {(2\pi/\tau(x))}^{1/2}\ N(y|y_0,1/\tau(x))\ \exp\{-c_j\tau(x)(x_j-x_0)^2/2 \}, $$
  where $N(y|m,v)$ is the p.d.f. of the univariate normal $N(m,v)$ evaluated at $y.$ 
\i The resulting BPDS p.d.f.  $f_j(y|x,\cM_j) \propto \alpha_j(y | x) p_j(y| x,\cM_j)$ is $N(m_j^*(x),v_j^*(x)) $
with $$m_j^*(x) = v_j^*(x)\{m_j +b_j x+\tau(x) v_j(x) y_0\}\quad\textrm{and}\quad v_j^*(x)=v_j(x)/\{1+\tau(x) v_j(x)\}.$$
\i The $a_j(x)$ are given by 
$$a_j(x)  =  {(2\pi/\tau(x))}^{1/2}  \exp\{-c_j\tau(x)( x_j-x_0)^2/2 \} 
 N(y_0|m_j + b_j x,  v_j(x)+1/\tau(x)).$$
 \ei

 Computations to define $\tau(x)$ at any given $x$-- based on a chosen target expected score $\starm$-- are easy via Newton-Raphson optimisation.  This relies on the identities  
$$ 
\frac{\partial a_j(x)}{\partial\tau(x)}= - \frac{a_j(x)}2 (r_j(x)+  d_j ) \quad\textrm{with}\quad r_j(x) = (m_j^*(x) - y_0)^2 + v_j^*(x)
\quad \textrm{and} \quad d_j = c_j (x_j-x_0)^2,$$  
and
$$
\frac{\partial^2 a_j(x)}{\partial\tau(x)^2} =  \frac{a_j(x)}4 \{ w_j(x) + 2  d_j r_j(x) + d_j^2 \}, $$
where $w_j(x) = (m_j^*(x) - y_0)^4 + 6 (m_j^*(x) - y_0)^2 v_j^*(x) + v_j^*(x)^2.$

BPDS analysis involves specifying the target (lower bound) on  the expected score. As noted above this will be anchored in the expected score under the initial mixture so that-- on the defined utility scale-- BPDS is defined as an extension of the initial mixture.  This example uses this concept as follows. Under the initial mixture, compute the model-specific expected scores at their optimal $x_j$ and then take the ET target score as the maximum; i.e.,  
$\starm = \max_{j=0{:}J} E_p[s_j(y,x_j)|\mathcal{M}_j].$ In the numerical example here, this yields 
$\starm = -0.046$. This is just one potential way of setting $\starm$, the portfolio example in Section ~\ref{sec:portfolioexample} provides another alternative.
Reanalysis with values slightly more or less extreme than this is also of interest, of course, while the following discussion adopts this specific value.

Under the reference BPDS utility of 
\eqn{DSutility}, the final BPDS  optimal decision $x$ maximises the expected utility in \eqn{DSoptimal}, i.e., 
\beas
\bar U(x) &=&  \tau(x) \sum_{j=\seq 0J} \tilde\pi_j(x) \int_y s_j(y,x)f_j(y|x,\cM_j)dy  \\ 
  &=&        - \tau(x)   \left\{ ( m_f(x)-y_0)^2 + v_f(x) + c_f(x) (x-x_0)^2 \right\}/2,
\eeas
where $m_f(x), v_f(x)$ are the mean and variance of $y$ under $f(y|x)$ and 
       $c_f(x) = \sum_{j=\seq 0J} \tilde\pi_j(x) c_j.$ 

\begin{figure}[htbp!]
    \centering
    \includegraphics[width=.6\textwidth]{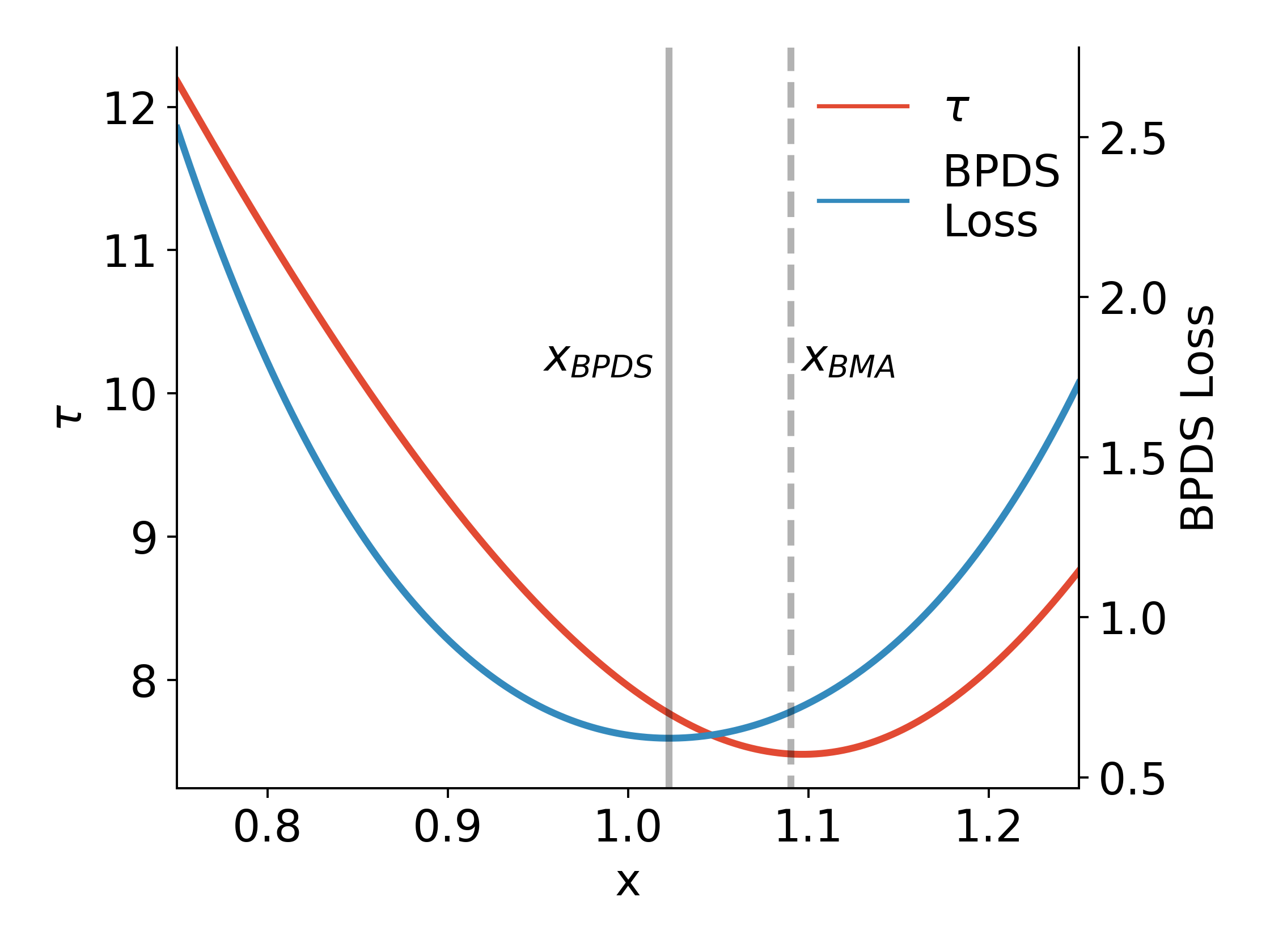}
    \caption{Optimal control example with $x_0=1$ and target $y_0=1$. ET--optimised $\tau(x)$ (left axis, blue line) and resulting expected BPDS utility (right axis, red line)  across a grid of $x$, with labelled $x_{BPDS} = 1.02$ that maximises the BPDS expected utility function, and $x_{BMA} = 1.09$. }
    \label{fig:ED_tau_util}
%
\bigskip\bigskip\bigskip\bigskip
%
   \centering
    \includegraphics[width=.6\textwidth]{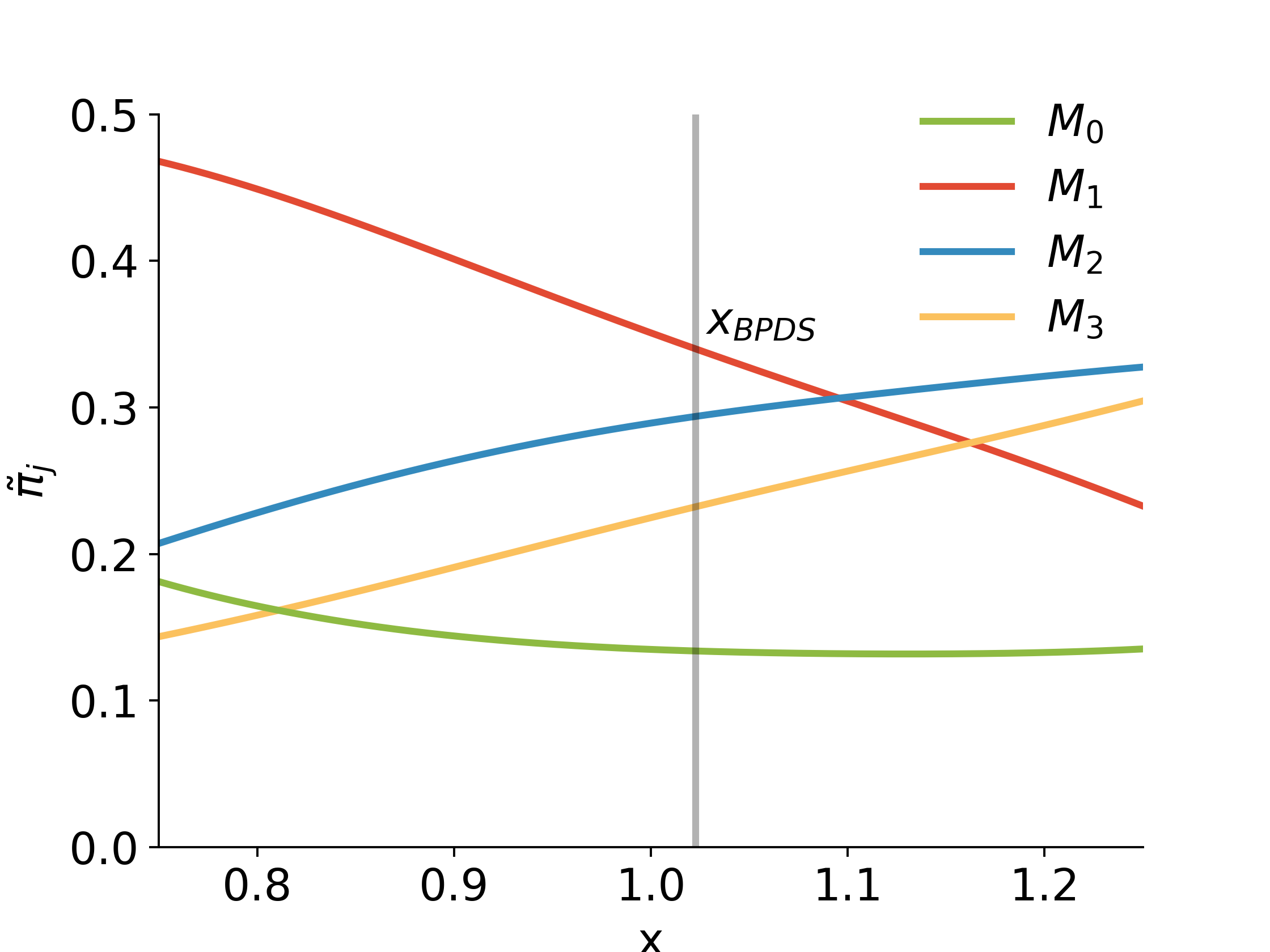}
    \caption{Optimal control example with $x_0=1$ and target $y_0=1$. Posterior $\tilde{\pi}_j(x)$  across a grid of values of $x$, with labelled $x_{BPDS} = 1.02$ that maximises the expected BPDS utility function.}
    \label{fig:ED_pi_tilde}
\end{figure}

It is easy to evaluate $\tau(x)$ and $\tilde{\pi}_j(x)$ and hence $\bar U(x)$ over a grid of $x$ values; see Figures~\ref{fig:ED_tau_util} and~\ref{fig:ED_pi_tilde}.  Observe the variation in the inferred ET parameter $\tau(x)$ as a function of $x,$ with a minimum of around 7. A comparison uses a proper BMA analysis. The BPDS optimal is $x_{BPDS}=1.02,$ a  lightly lower value than $x_{BMA} = 1.09$, shrunk a little more towards the current control setting $x_0=1.0$  The $\tilde{\pi}_j(x)$ shown in Figure~\ref{fig:ED_pi_tilde} vary substantially with $x$. By comparison, the BMA analysis assigns around $95\%$ probability on $\cM_2$ and almost all of the rest on $\cM_3.$   Compared to the uniform initial weights, BPDS weighs $\cM_1$ generally more highly and down-weights the baseline distribution-- a logical choice given the increased spread of the latter. $\cM_2$ and $\cM_3$  are more highly weighted as $x$ increases as their optimal $x_j$ values are higher. Table~\ref{tab:ED_results} shows the true synthetic resulting linear predictor $\hat y_j$  of $y$  and the implied loss under the optimal  $x_j$ of each $\cM_j.$  This loss  is evaluated as $L(x_j) = (\hat y_j-y_0)^2 + \beta^2 (x_j-x_0)^2$ at the underlying slope $\beta = 1.2$ of the synthetic data generating model.   BPDS achieves a significantly smaller loss than any other model, including the BMA approximation and the baseline.

\begin{table}[htpb!]
\centering
\begin{tabular}{lrrrr}
      \bf Model \ \ \ &  \ \ \ \bf Opt $x$ & \bf $\hat y$\ \ \  & \bf Loss & \ \ \bf \% Excess\\
\\
$\cM_1$ &   0.97 & 0.88 &  1.60 &     242 \\
$\cM_2$ &   1.09 & 1.02 &  1.14 &     143 \\
$\cM_3$ &   1.14 & 1.07 &  3.26 &     598 \\
baseline ($\cM_0$) &   1.06 & 0.99 &  0.61 &      30 \\
  Equal &   1.06 & 0.98 &  0.58 &      24 \\
    BMA &   1.09 & 1.02 &  1.21 &     158 \\
   BPDS &   1.02 & 0.94 &  0.47 &      --- \\
\end{tabular}
\bigskip
\caption{Optimal control example with $x_0=1$ and target $y_0=1$. 
For each model, decision $x$ and resulting linear predictor $\hat y$, the resulting empirical loss 
and implied percentage increase/excess relative to BPDS.   
Loss here is calculated as $100\{(\hat y-y_0)^2 + \beta^2(x-x_0)^2\}$ with true/synthetic model parameter $\beta = 1.2$.  BPDS dominates in terms of minimising loss, substantially so over BMA as well as over Equal weighting (the initial uniform $\pi_j$.) 
\label{tab:ED_results}
}
\end{table}

This analysis based on $c = 1$ equally penalises deviations from targets on the $x$ and $y$ scales. A repeat  analysis with $c=0.1$ provides expected results with $x_{BPDS}=1.10$ and resulting $y=1.03$, closer to $y_0$ and further from $x_0$ due to decreased weight on the $x-$dimension in the score/utility. BPDS further outperforms  BMA in that example with the latter having an over 620\% increase in loss relative to BPDS. BMA emphasises fit to the data and generally favours more conservative models, and in this case does put the majority of its weight on the model including all relevant variables. However, BMA completely eliminating the impact of $\cM_1$  leads to a selected control $x$ value that is too far away from $x_0$. BPDS operates under the assumption that the prior weights incorporate all past information about model accuracy, which in this example are uniform. $\cM_1$  is then given an increased weight under BPDS, since its forward-looking expectation generally indicates a higher score than the other models. The BPDS mixture is superior to the prior weighting: tilting towards models {\em expected} to perform well leads to an improved final decision.

\FloatBarrier

\section{Example: Dynamic Portfolio Investment Decisions \label{sec:portfolioexample}}

\subsection{Applied Setting, Data and Models} 

A second example concerns multivariate outcomes and 
time-varying models in sequential financial portfolio analysis. This involves time series of daily prices of $q=13$ assets including FX and market indices from August 2000 to the end of March 2010. On any day and based on the historical data and model analyses, the decision is to choose a portfolio weight vector to reallocate investments prior to moving to the next day. Ignoring time indices for clarity, $\y$ is the $1-$step ahead vector of percent returns on the $q$ assets, and the current decision is the $q-$vector of portfolio weights $\x$ such that the portfolio return is $\x'\y.$ Constraints include the sum-to-one condition $\x'\bone=1$ (no additions or withdrawals of capital are allowed, and all capital must be invested). 
In contrast to the design example, the predictive distributions $p_j(\y|\cdot)$ and BPDS weighting functions $\alpha_j(\cdot|\cdot)$ do not depend on the final portfolio  choice $\x$. The $\alpha_j(\cdot|\cdot)$ do, of course, depend on the $\cM_j$--specific optimal portfolio vectors $\x_j$.

Forecasting models $\cM_j$  are multivariate time-varying autoregressive (TV{--}VAR) dynamic linear models for log prices. Including the $t$ index for this discussion,  the model applies to  the vector   $\log(\mathbf{p}_{t})$ where $\mathbf{p}_{t}$ is the vector of asset prices on day $t$. These can be transformed to the return scale using $\y_{t} = \mathbf{p}_{t}/\mathbf{p}_{t-1} - 1$. TV{--}VAR models  include time-varying  covariance matrices addressing multivariate volatility~\citep[][chap.~9~\&~10] {Aguilar2000,IrieWest2018portfoliosBA,PradoFerreiraWest2021}.  A summary of the TV{--}VAR model form and specification is given in the Appendix here. 
All models have TV{--}VAR order 3, reflecting expected $2{-}3$ day momentum effects in price series: each univariate series $y_{t, i}$ is predicted by all of the values of  $\y_{t-3:t-1}$ with differing coefficients for each asset.  Each model uses a state evolution discount factor $\delta = 0.9995$. Models are distinguished through values of the volatility matrix discount factor $\beta$; this takes one of three values $\beta\in \{.94, .98, .995\}$, representing differing degrees of change in levels of volatility of each of the assets over time, as well as-- critically for portfolio analysis-- in the inter-dependencies among the assets as represented by time-varying covariances.  Technical details of the roles of $(\beta,\delta)$ are given in the Appendix.

In each TV{--}VAR model $\cM_j$, standard Bayesian forward-filtering analysis applies, and forecasting one-day ahead for local portfolio rebalancing (each day) uses Monte Carlo samples of the predictive distributions that map to the predictive mean vector and covariance matrix on the percent returns scale.   Within each model analysis, the model-specific decisions are based on quite standard Markowitz mean-variance portfolio optimisation: this simply minimises predicted portfolio variance among sum-to-one portfolios that share a daily expected return constrained to a specified target level $r_j$ (see, for example,~\citealp{PradoFerreiraWest2021},~section 10.4.7).  The set of $J=6$ BPDS model/decision pairs is defined by the use of two different target expected portfolio returns $r_j\in \{0.05,0.15\}$ coupled with each of the three TV{--}VAR models. 
For this example, the analyses are run \lq\lq blind'' and automated, with no interventions.  On each day, the resulting optimised portfolio weight vector $\x_j$ is  a vector of asset weights that has two constraints: they sum to one, i.e., $\x_j'\bone=1$, and have the specified target mean property $E(\x_j'\y|\cM_j) = r_j$.

\begin{figure}[t!]
    \centering
    \includegraphics[width=.95\textwidth]{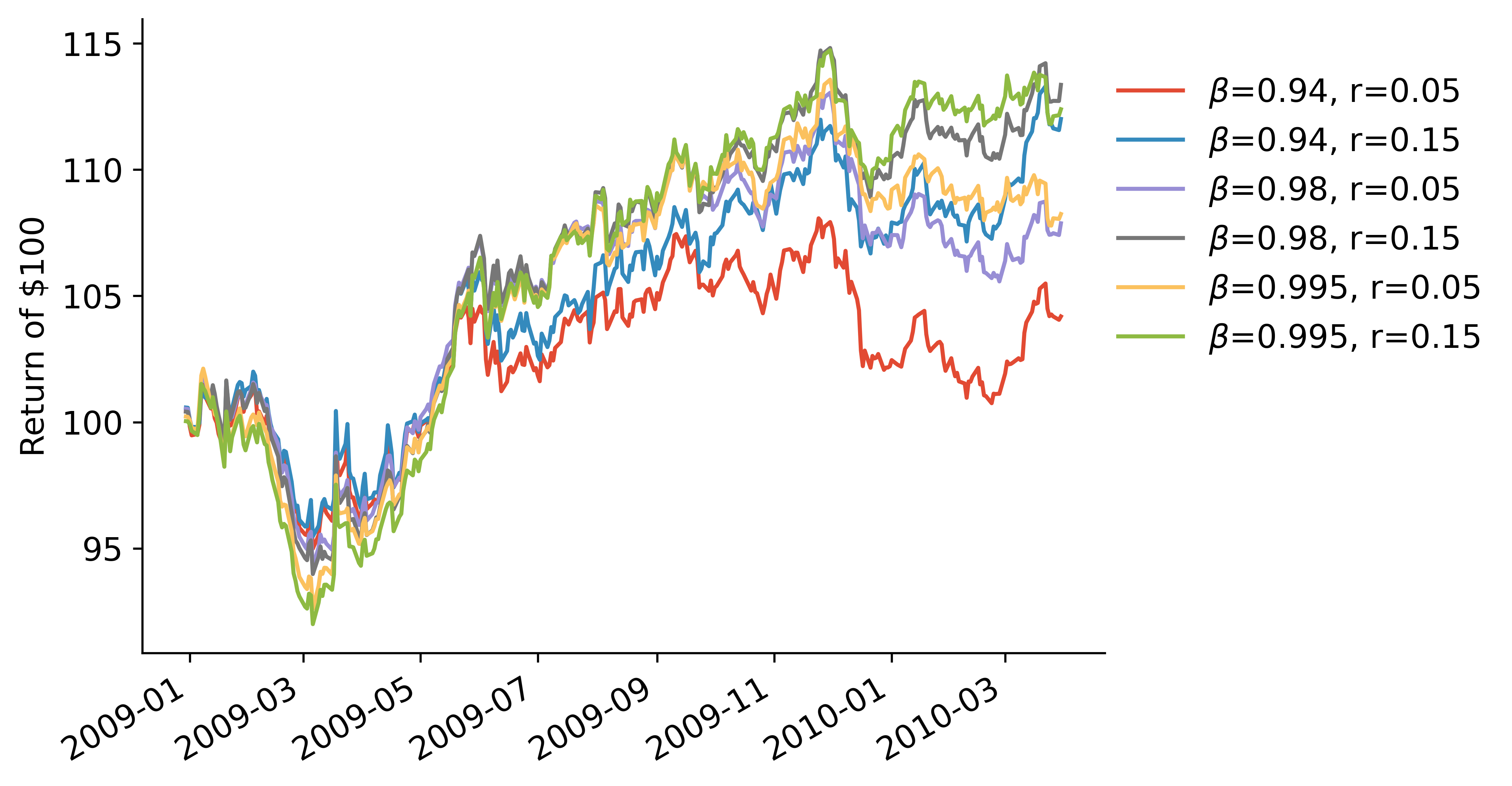}
    \caption{Returns on \$100 invested from January 2009 to the end of March 2010 for $J=6$ TV{--}VAR models with differing volatility matrix discount factors $\beta,$ evaluated using Markowitz portfolios with two different target returns $r_j.$}
    \label{fig:portfolio_TVAR}
\end{figure}

The period of 2{,}195 trading days to the end of 2008 provide model training data,  and the following 325 trading  days
to the end of March 2010 define the test period.   Figure~\ref{fig:portfolio_TVAR} shows the model-specific portfolio returns over the test period. Initial losses as national economies exit the very volatile recessionary period are followed by market stabilisation and a period of generally positive returns.  Note the eventual superiority of the more stable volatility models (i.e., larger volatility discount factor $\beta$) as markets stabilise and recover from the recessionary period. This provides a good setting to evaluate BPDS, as it will ideally show more resilience to  the initial downturn while taking advantage of   increasing returns once model predictions and decision-outcome performance improves.
 
\FloatBarrier

\subsection{BPDS Utility-based Score \label{sec:BPDSportfolioutility}} 

BPDS analysis adopts the bivariate score vector 
\beq{port_score}\s_j(\y, \x_j) = [ \,\x_j'\y,\, -(\x_j'\y-m)^2/2 \,]'\eeq
for a chosen target return $\portar=0.1$, notably the midpoint of the model-specific potential targets. This is motivated by
the classic,  bounded-above risk-averse exponential utility   function for return $r$ given by 
$V_e(r) = -\exp (-r/d)$ for some risk tolerance level $d>0$. A second-order Taylor series approximation around $r=\portar$ gives $V_e(r)\approx V_q(r) = V_e(r)\{1-(r-\portar)/d+(r-\portar)^2/2d^2\}.$ This is a very accurate approximation to the bounded exponential utility function across ranges of $r,\portar$ which correspond to realistic models for return prediction in portfolio analyses, especially for daily returns.
Note that $V_q(r)$ is maximised at $r=\portar+d$. Using a univariate score $s_j(\y, \x_j) = \x_j'\y -(\x_j'\y-\portar)^2/2d$ gives $\alpha_j(\y, \x_j) = \exp\{\tau\x_j'\y -\tau(\x_j'\y-\portar)^2/2d\}$.  However, using the bivariate score allows flexibility in defining the levels of risk tolerance 
as $d$ is implicitly specified; the
bivariate score gives  $\alpha_j(\y, \x_j) = \exp\{\tau_1\x_j'\y -\tau_2(\x_j'\y-\portar)^2\}/2$ which agrees with the univariate case if/when  $\tau = \tau_1$ and $d= \tau_1/\tau_2$, and thus defines a maximising target return  $\portar + d$ at this level of risk tolerance. The bivariate score allows more flexibility in the role of the return/risk elements and their impact in the BPDS analysis,  while overlaying and replicating what the classical exponential utility univariate score defines as a special case. 

\subsection{Sequential Analysis and Operational BPDS for Time Series Portfolios} 

Recognise the sequential time series and decision setting by extending the notation to add a time subscript $t.$ In the example, this indexes working days.   At any current time $t$ 
the model set,  predictive distributions, initial model probabilities, and BPDS score vectors and targets are now all indexed by $t$ as well as model index $j;$ for example, model probabilities  $\pi_{t,j}$ define weights that incorporate historical performance on all data and aspects of the analysis up to but not including time $t.$  At each time point, the BPDS analysis will take as inputs the current model summaries and define a time-specific ET vector $\btau_t$ for the resulting BPDS mixture predictive and decisions. As $t$ evolves, the analysis will repeat.

\subsubsection{Time-Specific Initial BPDS Probabilities and Baseline} 

The analysis builds on and  extends prior developments of Bayesian model weighting based on historical predictive performance using AVS~\citep{LavineLindonWest2021avs}.  For each time $t,$  the current initial model probabilities are taken as $\pi_{t,j}\propto (\pi_{t-1,j})^{\gamma}\exp\{\btau_{t-1}' \s_{t-1,j}(\y_{t-1}, \x_{t-1,j})\}$ where the AVS model discount factor $0{\ll}\gamma{<}1$ down-weights  the impact of model predictive performance based on more distant past data~\citep{ZhaoXieWest2016ASMBI}; the example here takes $\gamma = 0.95$. This is a coherent extension of AVS to form an extended BPDS analysis in which the forward-looking BPDS score functions are also used as realised utilities in weighting historical model performance.  Time $t=0$ model probabilities $\pi_{0,j}$ are taken as uniform.  The vector score is as above in \eqn{port_score},  now indexed by $t$ as well as $j$ to make explicit the variation in time.    

The time $t$ initial mixture model defined by these AVS weights and densities $p_{t,j}(\cdot|\cdot)$ is referred to as the \lq\lq AVS\rq\rq\ model.  The final aspect of model structure is the baseline. For each $t,$  each of the TV{--}VAR models generates a theoretical one-step ahead forecast distribution for log asset prices that is multivariate T, with parameters that are model-specific and updated in the standard forward-filtering analysis.  The BPDS baseline $p_{t,0}(\cdot|\cdot)$ is taken as 
a T distribution with 9 degrees of freedom, the same location as the AVS mixture, and variance matrix of that of the AVS mixture inflated by $1/\delta$ with $\delta=0.135$. Consistent with the earlier discussed desiderata for the choice of baseline, this represents a neutral-but-diffuse predictive distribution relative to the current set of models in the AVS mixture. The
baseline initial probability $\pi_{t,0}$ is also defined via the AVS  weighting,  extending the analysis to include $j=0$ giving $j=\seq 0J$ dynamic models. The set of model predictive densities for time $t$ is denoted by  $\cP_t  = \{ p_{t,0}(\y_t|\cM_{t, 0}),\ldots,p_{t,J}(\y_t|\cM_{t,J}) \}.$
    
\subsubsection{BPDS Decision Focus, Constraints and Implementation} 

The target score at any time point is set adaptively based on realistic assessment of expected returns and risk levels using the \lq\lq current'' set of models at each time.  The analysis adopts $\starbm_t = (1.05 \bar{\starm}_{t,1}, 0.9 \bar{\starm}_{t,2})'$ as the target score at time $t$ where the $\bar{\starm}_\ast$ elements are the expected scores under the current, time $t$ mixture over the models.  This aims for a 5\% greater expected return and a 10\% decrease in the squared deviation from target return $\portar=0.1$ on each day. 
Further, this example setting is one in which restrictions on elements of the now time-dependent $\btau_t$ are desirable.
The elements of the score vector are obviously dependent, and it is relevant to constrain $\btau_t$ to have positive elements so that higher expected returns and lower expected risk are rewarded. The ratio $d_t=\tau_{t,1}/\tau_{t,2}$ has the interpretation of a risk aversion parameter through the connection to  exponential utility,  and a smaller value is  desirable; here the additional constraints $0<\tau_{t,1}<0.1\tau_{t,2}$ address these desiderata and additionally constrains the ET optimisation analysis for $\btau_t$ at each time point.   
Note also that there is no analytic form for the  $a_{t,j}$ terms in this setting;  Monte Carlo analysis is used to evaluate the (time $t$ dependent extensions of) integrals in~\eqn{jderivs} within numerical optimisation to solve for $\btau_t$. 
    
\subsubsection{BPDS Optimal Portfolio Decisions} 

With the time dependence explicit in notation,  $f_t(\y_t|\cP_t)$ is now the  one-step ahead BPDS mixture predictive density for returns from time $t-1$ to time $t.$ The operational portfolio optimisation adopts the reference utility as in Section~\ref{sec:finaldecision}, with the extension to time-dependence here.  In this example with linear and quadratic elements of $\s_{t,j}(\y_t, \x_t)$, 
the expected reference utility function from \eqn{DSoptimal} at time $t$ reduces to 
 \beq{port_exp_score_time_t}
        \bar U_t(\x_t)  =  \tau_{t,1} \x_t'\f_t -\tau_{t,2}(\x_t'\f_t - \portar)^2/2 - \tau_{t,2} \x_t'\V_t \x_t/2
    \eeq
where $\f_t$ and $\V_t$ are the forecast mean vector and variance matrix of $f_t(\y_t|\cP_t)$.   
    
This is an example where additional constraints on the final decision vector $\x_t$ are relevant. First, add the usual unit-sum constraint  $\bone'\x_t = 1.$
Second, constrain portfolio decisions such that $\x_t'\f_t = \portar_t^*$ for some specified target $\portar_t^*.$  Subject to these constraints,  maximising \eqn{port_exp_score_time_t} reduces to the Markowitz solution with target expected return, with a simple analytic solution for $\x_t$.  The analysis here takes $\portar_t^* =\portar+ d_t$ with risk tolerance $d_t = \tau_{t, 1}/\tau_{t,2}$; this maximises expected utility  as noted earlier in Section~\ref{sec:BPDSportfolioutility}.  With earlier discussed settings this implies that 
$0 .1 = \portar  < \portar_t^*< 0.2$, allowing for potential time-dependent improvements over the original target $\portar $. 

Comparisons are with BMA--based model averaging and AVS. In each,  the analysis applies Markowitz optimisation  with target expected 
 return $\sum_{j=\seq 1J} \pi_{t,j} r_{t,j}$ where the $\pi_{t,j}$ are the initial mixture model weights and $r_{t,j}$ is expected return in $\cM_j$ at time $t$. Note that BMA  weights models  wholly based on predictive fit to the data scored by the sequences of density values $p_{t, j}(\y_{t}| \mathcal{M}_{t, j})$, and does not take any account of the decision focus the models are designed to address.   Thus, BMA cannot distinguish between models with identical discount factors but different expected returns $r_{t,j}$,  so the portfolio weights will always evenly split weight between the two comparison targets $r_{t,j}$, giving an expected return of $0.1$. In contrast, AVS weighting accounts for the $\x_{t,j}$ and historical return outcomes in each model, and so differentially weights models with the same discount factor;   AVS expected returns then vary over time as, of course, do those from BPDS.

\subsection{Some Summary Results} 

\begin{figure}[p!]
    \centering
    \includegraphics[width=.8\textwidth]{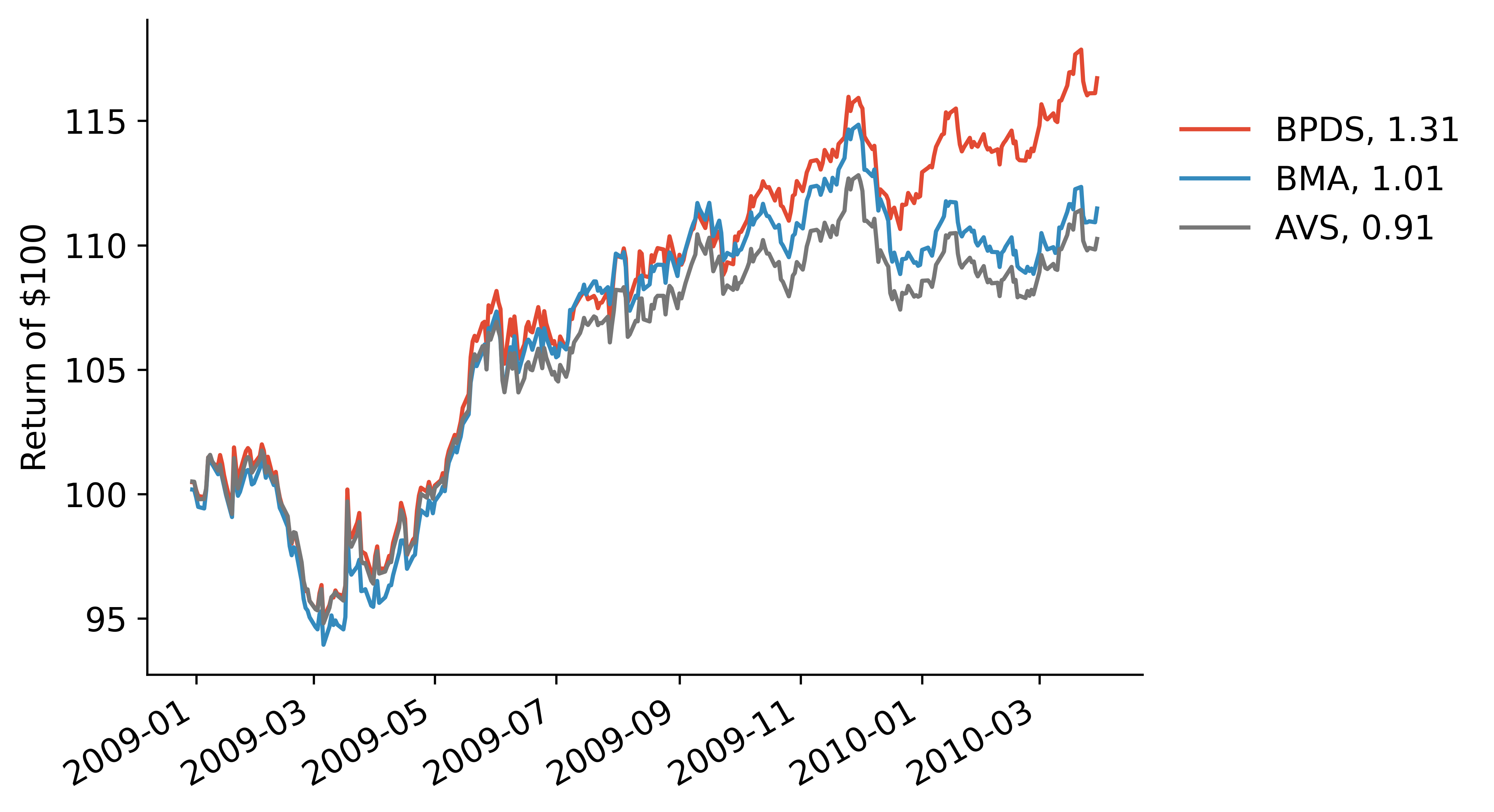}
    \caption{Returns on \$100 invested from January 2009 to the end of March 2010 evaluated in each of the BPDS, BMA and AVS analyses, as well as their Sharpe ratio (see the legend, formatted as name, Sharpe). The Sharpe ratio is calculated as the annualised ratio of mean return to standard deviation of return.}
    \label{fig:portfolio_returns}
\bigskip\bigskip
    \includegraphics[width=.7\textwidth]{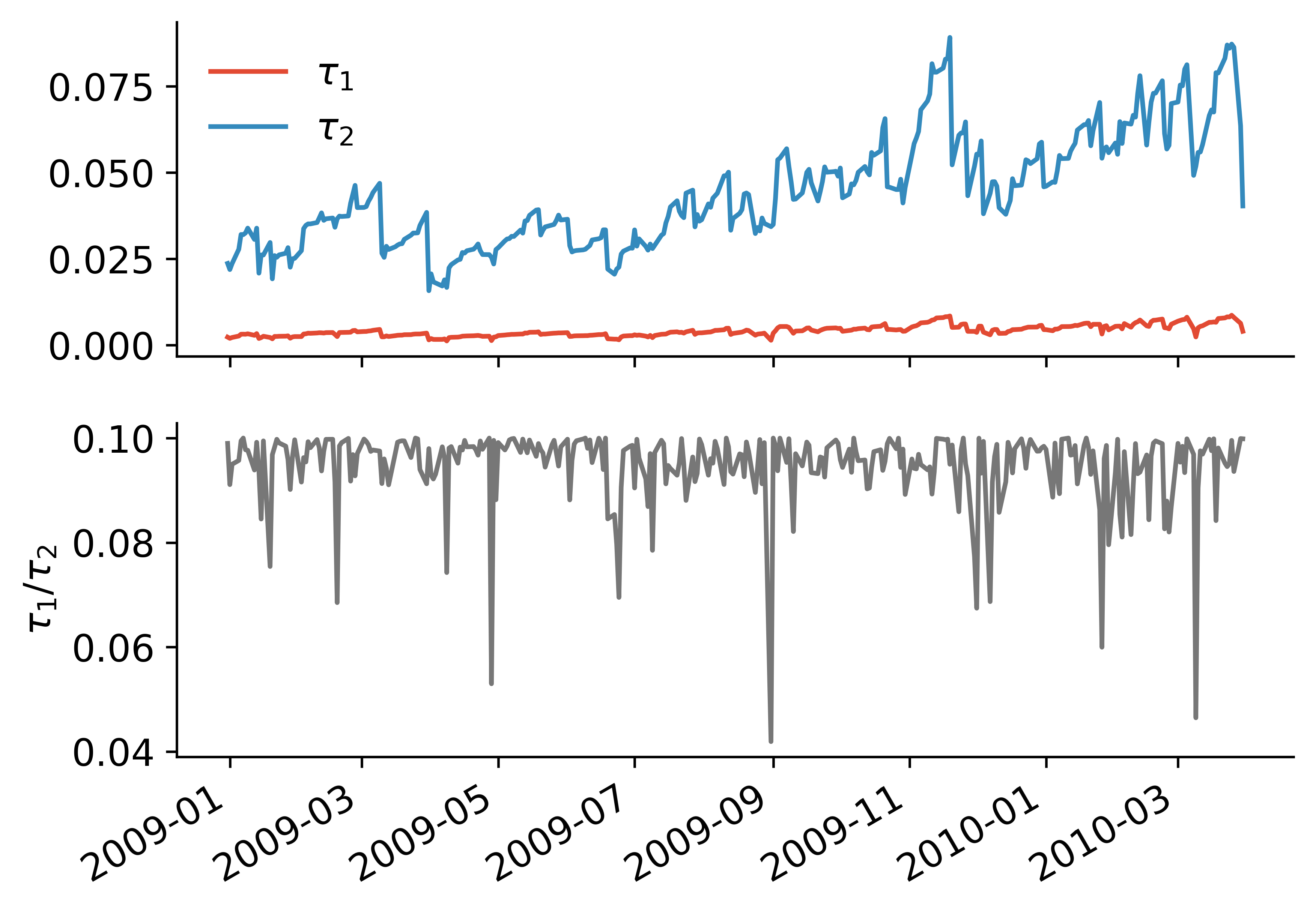}
    \caption{Resulting $\btau_t$ (upper) and ratio $\tau_{t,1}/\tau_{t,2}$ (lower) from the BPDS portfolio analysis. The latter shows that the target expected return 
     $m + \tau_{t,1}/\tau_{t,2}$ in the BPDS analysis exhibits meaningful levels of  variation throughout the time period. }
    \label{fig:portfolio_tau}
\end{figure}

For each of the three analyses-- based on BPDS, BMA, and AVS, respectively-- cumulative returns and empirical Sharpe ratios  are shown in Figure~\ref{fig:portfolio_returns}. At the start of the test period, all models perform somewhat poorly through the  tail-end of the recession.  BPDS shows a slight advantage in this time period, but cannot completely avoid the downturn effects. However, once the economy begins to recover and information flows stabilise, BPDS is able to achieve a significantly higher return in comparison to BMA or AVS. Interestingly, basic AVS performs worse than BMA overall, which is essentially due to constraining the AVS tilting vector $\btau$ to have smaller values;  this is necessary in the BPDS analysis and for comparison, but leads to lack of differentially weighting \lq\lq good'' historical performance in basic AVS.   Sharpe ratio summaries support similar  conclusions. BPDS ultimately dominates both BMA and AVS. BPDS achieves increased returns without corresponding increased risk, exemplifying the prospects for prediction and decision improvements. 
 
 \begin{figure}[p!]
    \centering
    \includegraphics[width=.85\textwidth]{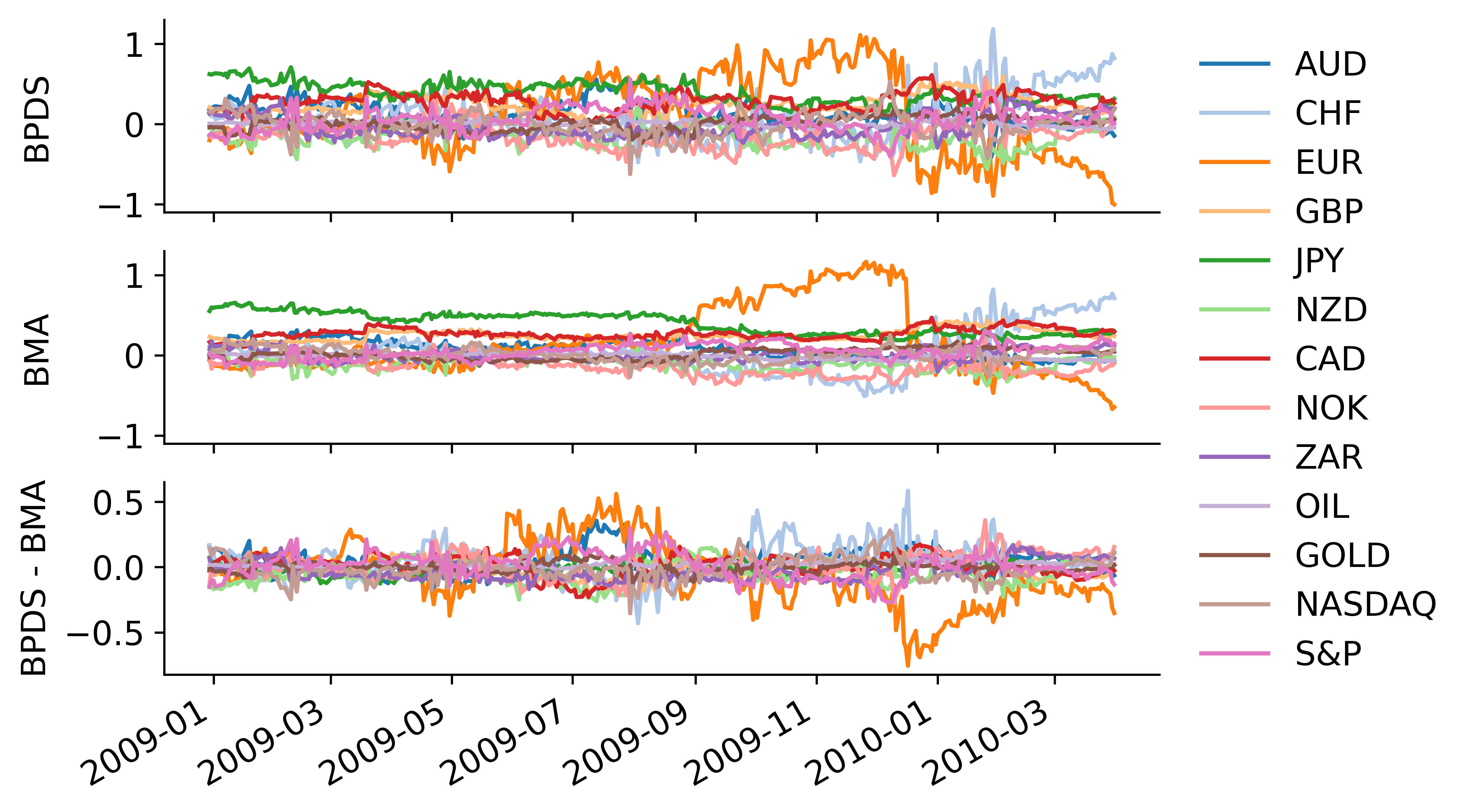}
    \caption{Time trajectories of optimised portfolio weights for each of the 13 assets in analyses based on BPDS (upper) and BMA (middle) models, and the difference of weights-- BPDS minus BMA (lower).}
    \label{fig:portfolio_weights}
\bigskip\bigskip
    \includegraphics[width=.95\textwidth]{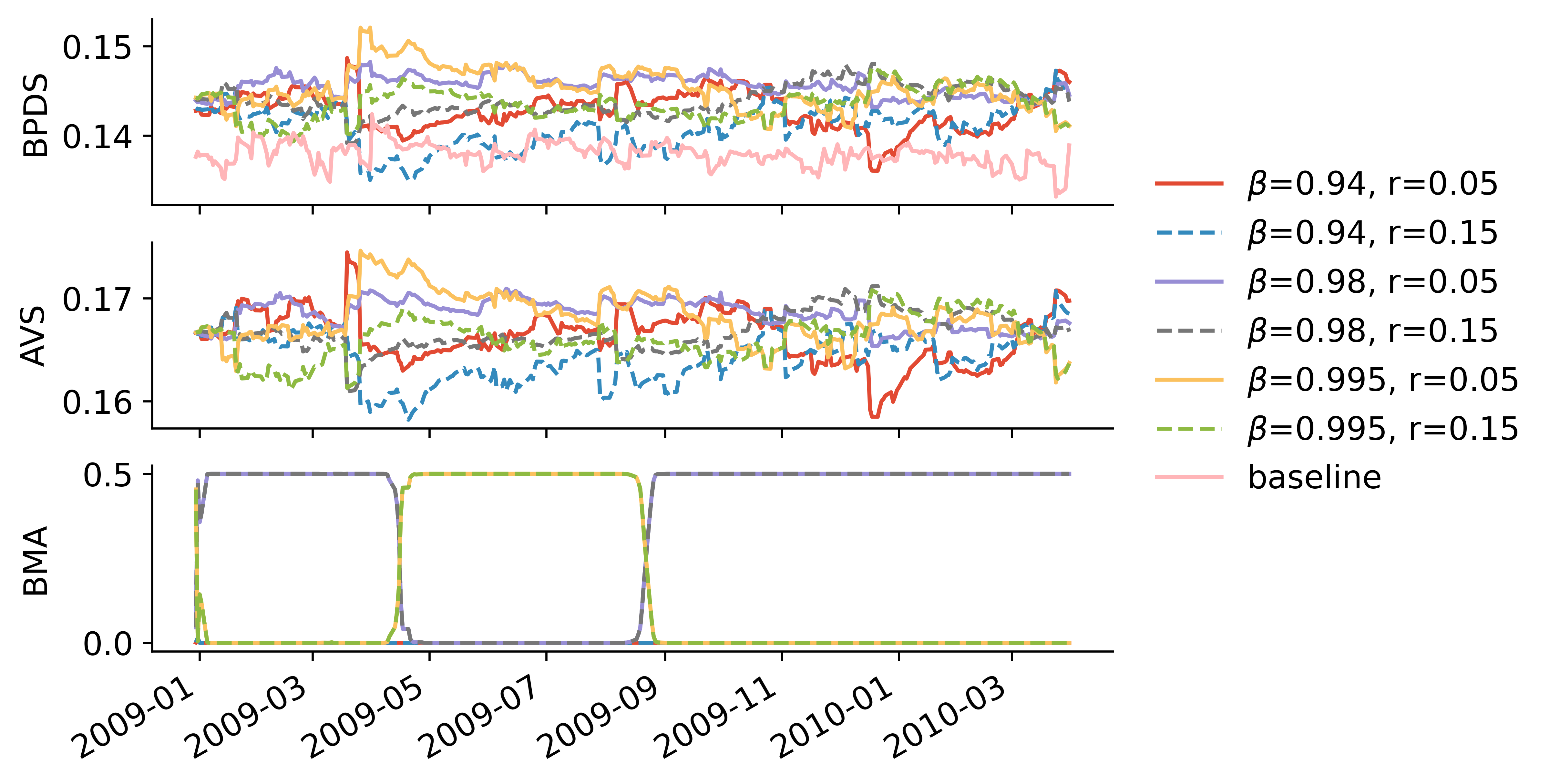}
    \caption{Model weights $\pi^*_j$ for BPDS (upper), AVS (middle) and BMA (lower). The baseline model is only included in the BPDS analysis by definition.}
    \label{fig:portfolio_pis}
\end{figure}

Time trajectories of the optimised BPDS $\btau_t$ vectors exhibit more variation in weightings on the second score component than the first; see Figure~\ref{fig:portfolio_tau}. 
This is partly due to the scale, as $\tau_{t,1}$ is constrained to be smaller than $\tau_{t,2}$. There are also influences from the model-specific portfolio constraints which target somewhat similar expected returns across models over the time period. That is, there is limited variation over time in expected returns (the first score vector elements)  but substantially more diversity in terms of forecast uncertainty about returns (related to the second score vector element). The ratio of the two elements defines the target return for the BPDS analysis, which then does vary substantially over the time period while generally staying close to the  defined $0.1$ boundary.  This leads to BPDS portfolio weights being somewhat more adaptive than those from the BMA analysis (see Figure~\ref{fig:portfolio_weights}),   while indicating that the BPDS foundation with a differing target score generally leads to  higher realised scores and increased returns.

Figure~\ref{fig:portfolio_weights}  displays trajectories of  sequentially revised portfolio weights from the BPDS and BMA analyses, and the difference in weights, BPDS minus BMA. The optimised weights have similar patterns over time, while there is generally more variation under BPDS. This decision-guided difference in the BPDS analysis partly drives small improvements in portfolio outcomes across the time period. More noticeable differences appear in mid-2009 and into 2010, particularly with respect to the weights on the Euro and, to a lesser degree, the Swiss franc.  Here BPDS is more adaptive to the increasingly volatile period for the Euro linked  to the developing Euro-zone sovereign debt crisis. Relative to BMA, the BPDS analysis is able to capitalise on this increased volatility and, in later 2009, begins to more aggressively short investments in the Euro. This later position is a key driver of the more evident increased returns seen in Figure~\ref{fig:portfolio_returns}.

Trajectories of the BPDS, BMA and AVS model probabilities are shown  in Figure~\ref{fig:portfolio_pis}. A key feature is the typical performance of BMA. The BMA probabilities  switch dramatically between favouring the two related models with $\beta=0.98$ and the two with $\beta=0.995,$ while the most adaptive models with $\beta=0.94$ receive negligible weight. The latter point is a key issue and concern in terms of time adaptability in dynamic modelling. 
On BPDS, note that the baseline is generally down-weighted, including during the initial period where the set of models performs relatively poorly overall. The baseline is relatively disadvantaged due to its inflated spread without a significant advantage in location over this period.     
Then, BPDS initially down-weights the more adaptive models in comparison to AVS due to the variance penalty in the score function rewarding more stable models.  BPDS and AVS favour more stable models towards the end of the testing period as information stabilises.  BPDS weights-- while slightly more volatile and adaptive locally-- tend to shrink AVS weights, inducing more robustness in part due to the role of the baseline.

\FloatBarrier 
 
\section{Summary Comments\label{sec:conclude}}

The  framework of Bayesian predictive decision synthesis expands traditional thinking about evaluating, comparing and combining models to explicitly integrate decision-analytic goals.   Growing from a theoretical foundation in Bayesian predictive synthesis and empirical approaches to goal-focused model uncertainty analysis, BPDS defines a theoretical basis for explicitly recognising both predictive and decision goals, integrating the 
intended uses of models into the model uncertainty and combination framework. Recognising and \lq\lq rewarding'' models through utility functions that {\em anticipate} improved decision outcomes together with predictions is the central concept.  The integration with focused predictive methods, such as AVS,  unites the core ideas of model scoring on past data with \lq\lq outcome dependent'', anticipatory model weightings to define novel methodology.  The use of entropic tilting is both foundational and enabling in implementation; ET also aids the interpretation of BPDS, important in bringing the approach to new applications. 

The detailed examples highlight core aspects of the methodology and demonstrate the potential for BPDS to yield improvements in Bayesian decision analysis in two very different settings.   The examples highlight the flexibility of BPDS to address a variety of prediction and decision problems, and its customisation via the use of context-dependent utility functions to define BPDS scores relevant for each applied setting. They also demonstrate that BPDS can generate  interpretable outputs which may improve understanding of models differences in the decision context, in addition to generating potentially superior outcomes relative to standard model combination approaches.  

The examples also note some basic implementation and computational aspects of the methodology. The use of coupled simulation and optimisation methods is generic and will be the basis for implementation in broader and larger-scale problems.  BPDS predictive distributions will typically be simulated using samples from the prior mixture, delivering direct Monte Carlo approximations to integrals required in iterative solution of the optimisation problems to maximise expected utilities subject to ET and perhaps other constraints. Our design example is simple and expository; there the evaluation of decision variables across a discrete grid is easy computationally and has a convex and unique solution.  In more realistic, applied design settings-- such as in the use of Gaussian process regression and other reinforcement learning contexts-- decision variables are multivariate, chosen BPDS utility functions may not lead to convex forms for solution,   and additional constraints may be desired as in the portfolio example here.    In terms of scalability of computations,  note that second-order gradient methods for optimisation to define tilting vectors scale requires matrix inversions that scale cubically in the dimension of the score vectors. Alternative numerical search approaches that scale linearly might be of interest in settings models with higher-dimensional scores.  
  
The use of AVS-style prior weights on models in the portfolio example is a setting in which the BPDS predictive distributions do not depend on the pending decision variables.  Extension of AVS-style weights to settings in which this dependence exists and is key-- such as in more elaborate developments of the basic optimal design example-- is an open question.   As with BPS, the new BPDS framework admits the use of a baseline distribution that requires specification in any application. Some desiderata are noted in Section 2.2.3 above;  theoretical study of the impact of any chosen baseline represents an area for future research (relevant more broadly than BPDS).  There is also theoretical interest in more foundational aspects of ET in the BPDS framework. This includes questions about the relationships between target score vectors and entropic tilting parameter vectors,  delineation of problems in which ET has no solution as a result of specific choices of target scores,  the roles of bounded utility functions to define scores, and theoretical implications of imposing additional constraints on the tilting parameters.  On the question of choice of target scores, large deviations from that under the initial mixture will represent discarding much of the information and experience the latter is based on, so is obviously not of interest. This also relates to the question of achievable solutions for any chosen target score; while this is a matter for theoretical study,  the BPDS analysis analysis can (sometimes)   identify  infeasible choices of target expected scores via failure of the numerical optimisation to identify implied tilting vectors. 
BPDS is open to including arbitrary choices of utility functions to define score vectors.  The current examples involve context-specific scores, while other applications could involve more general score choices-- such as scores related to aspects of probabilistic calibration~\citep[e.g.][]{GneitingRaftery2007}. 
Then, links with other areas of decision analysis involving multi-dimensional utilities are of interest; BPDS score vectors can reflect multiple aspects of a decision problem with the potential for conflict between dimensions.   In sequential time series decision problems, as exemplified by the financial portfolio example, there are a number of open questions.  Extended or alternative choices of score functions that reflect risk:reward in different ways, turnover in portfolio allocations,  multi-step-ahead portfolio construction, and psychological as well as financial dimensions~\citep[e.g.][]{IrieWest2018portfoliosBA}, are some general topics of interest. 

Extensions of BPDS model probability weighting functions are of interest to address questions of relationships and dependencies across the set of models in the mixture.  This has been a theme in recent developments of Bayesian predictive synthesis in purely predictive time series applications~\citep{McAlinnEtAldiscussionBA2018,McAlinnWest2018}. The concept will apply to BPDS by extending mixture-model based BPS to allow 
weights on model $\cM_j$ to depend on aspects of predictive and decision performance of other models $\cM_i$ for $i\ne j$.
There is potential practical value in this; for example, a subset of models generating very similar predictions and decisions might then each have appropriately lesser roles in the ET synthesis, accounting for the so-called herding effect~\citep[e.g.][]{JohnsonWest2022}.   There are also opportunities to explore more probabilistic choices of BPDS score functions, including score vector elements that relate to aspects of probability forecast calibration and proper scoring rules~\citep[e.g.][]{RafteryEtAl2005}.
These and other extensions are open for future research.

\newpage

\section*{Appendix: Multivariate TV{--}VAR Dynamic Models\label{sec:MV-DLMs}}

Summaries of standard forecasting models used in the portfolio example are given here. All material is sourced from~\citealp[][section 10.8]{PradoFerreiraWest2021};   notation (including that for Wishart, inverse-Wishart, matrix normal and matrix normal-inverse Wishart distributions) is as detailed there. 

Over $t=1,2,\ldots $ the 
$(q\times 1)-$vector time series $\y_t$  follows a random walk multivariate DLM   
$$
    \y_t'= \F_t' \bTheta_t + \bnu_t', \ \ \bnu_t\sim N(\bzero, \V_t), \qquad\textrm{and}\qquad 
    \bTheta_t = \bTheta_{t-1} + \boldsymbol{\Omega}_t, \ \ \boldsymbol{\Omega}_t \sim N(\bzero, \mathbf{W}_t, \V_t),
$$
with components as follows. At time $t$: 
\begin{itemize} \itemsep-5pt
\item  $\bTheta_t$ is the $(p\times q)$ time-varying state matrix, and $\V_t$ is  the $(q\times q)$ time-varying volatility matrix.
\item  $\bnu_t$ is the $(q\times 1)$ observation error vector, and $\bOmega_t$ is the $(p\times q)$ state evolution   error matrix.
\item  $\F_t$ is the $(p\times 1)$  regression vector of known constants and/or predictors.
\item  $\W_t$ is the known ($p\times p$) state evolution covariance matrix.
\end{itemize}
Write $\cD_t$ for all past data and information prior to time $t.$ Initialise analysis using the time $t=0$ distribution
 $(\bTheta_0, \V_0 |\cD_0) \sim NIW(\M_0, \C_0, h_{0}-q+1, \D_0)$. Sequential learning and forecasting over time follows a conjugate analysis structure with, for all $t= 1,2,\ldots $ the following summaries:
 \begin{itemize} \itemsep-5pt
\item  Time $t-1$ posterior  $(\bTheta_{t-1}, \V_{t-1} |\cD_{t-1}) \sim NIW(\M_{t-1}, \C_{t-1}, h_{t-1}-q+1, \D_{t-1})$.
\item  Time $t$ prior 
 $(\bTheta_t, \V_t |\cD_{t-1}) \sim NIW(\M_{t-1}, \R_t, \beta h_{t-1}-q+1, \beta \D_{t-1})$ with $\R_t = \C_{t-1}+\W_t$. The elements $\beta,\W_t$ here are defined below. 
\item   $1-$step ahead forecast distribution $(\y_t|\cD_{t-1}) \sim T_{\beta h_{t-1}}(\f_t, c_t\D_{t-1}) $ with $\f_t =\M_{t-1}'\F_t$ and $c_t = \beta q_t/(\beta h_{t-1}-q+1).$ 
\item  Time $t$ posterior update on observing $\y_t$ gives $(\bTheta_t, \V_t |\cD_t) \sim NIW(\M_t, \C_t, h_t-q+1, \D_t)$ with update equations  

\phantom{.}\quad $
\M_t = \M_{t-1} + \A_t \e_t', \quad  \C_t = \R_t - q_t\A_t\A_t', \quad 
 h_t = \beta h_{t-1}+1\  \textrm{and} \  \D_t = \beta \D_{t-1} + \e_t\e_t'/q_t$
 
\noindent where  $\e_t = \y_t- \f_t$ and $ \A_t =\R_t\F_t/q_t $ with $q_t = 1+\F_t'\R_t\F_t.$
\end{itemize}
The extent of time variation  in the state and volatility matrices is controlled by discount factors $(\delta,\beta),$ each in $(0,1]$ and typically closer to 1. First,  the state evolution  covariance matrix is 
$\W_t = \C_{t-1}  (1-\delta)/\delta$ based on the {\em state discount factor} $\delta.$  This simplifies  details above as it implies 
$\R_t = \C_{t-1}/\delta.$  Second, the volatility matrix  $\V_t$ follows a dynamic inverse Wishart evolution depending on the {\em volatility discount factor} $\beta$; this is evident above in the discounting of the Wishart degrees of freedom and scale matrix in the evolution from time $t-1$ to $t.$

\subsection*{TV{--}VAR Models}
Time-varying, vector autoregressive (TV{--}VAR) models are special cases.  A TV{--}VAR model of order $r$ is defined by the specification 
$\F_t' = ( 1, \y_{t-1}',  \y_{t-2}', \ldots, \y_{t-r}').$  Hence each univariate time series element of $\y_t$ is regression on recent past values of all $q$ univariate series up to and including lag $r$, together with an intercept term. The $i^{th}$ column of the state matrix $\bTheta_t$ then has the time-varying 
intercept and regression coefficients on these predictors for univariate series $i$. 
 
\section*{Supplementary Material} 

\cite{TallmanWestET2022} is a supplementary document  providing detailed background and theoretical developments of entropic tilting and related topics. This details the  theory of ET for conditioning predictive distributions on constraints, includes new results related to connections with regular exponential families of distributions, elaborates on the relaxed entropic tilting, and provides a number of examples. Supplementary code to replicate the design example is available and will be provided in the final version.

\section*{Acknowledgements}
The research reported here was partially supported by the National Science Foundation through the NSF Graduate Research
Fellowship Program grant DGE~2139754,  and by $84.51^\circ$ Labs.  Any opinions, findings, and conclusions or recommendations expressed in this material are those of the authors and do not necessarily reflect
the views of the National Science Foundation or the views of $84.51^\circ$. The authors acknowledge useful discussions with Christoph Hellmayr $(84.51^\circ$~Labs.),  Joseph Lawson and Graham Tierney (Department of Statistical Science, Duke University), Kaoru Irie (University of Tokyo), Jouchi Nakajima (Institute of Economic Research, Hitotsubashi University), with Gary Koop (University of Strathclyde) and Tony Chernis (Bank of Canada), and with several participants at the 2022 World Meeting of the International Society for Bayesian Analysis (Montreal, July 2022). 
 

\bibliography{BPDSreferences}
\normalsize
\bibliographystyle{chicago}

\end{document}